\documentclass[twocolumn,english,journal, final]{IEEEtran}
\usepackage[T1]{fontenc}
\usepackage{geometry}
\usepackage{float}
\usepackage{units}
\usepackage{mathtools}
\usepackage{amsmath}
\usepackage{amssymb}
\usepackage{graphicx}
\usepackage{babel}

\makeatletter

\providecommand{\tabularnewline}{\\}
\floatstyle{ruled}
\newfloat{algorithm}{tbp}{loa}
\providecommand{\algorithmname}{Algorithm}
\floatname{algorithm}{\protect\algorithmname}

\usepackage{babel}
\usepackage[caption=false,font=footnotesize]{subfig}
\usepackage{balance}

\usepackage{tikz}
\usepackage{algorithm,algpseudocode}

\usepackage{pgfplots}
\usepgfplotslibrary{groupplots}
\usepackage{graphicx}
\pgfplotsset{compat=newest}
\usepackage{units}
\usepackage{amsmath, amsbsy, amssymb}
\usepackage{tikzscale}
\usepackage{color}
\usetikzlibrary{plotmarks}
\usetikzlibrary{dsp,chains}
\usetikzlibrary{spy}
\usetikzlibrary{fadings}
\usetikzlibrary{positioning,chains,fit,shapes,calc}

\pgfplotsset{
colormap={jetlight}{rgb = (  1.00000000,   1.00000000,   1.00000000),rgb = (  0.99607843,   0.99607843,   0.99816176),rgb = (  0.99215686,   0.99215686,   0.99644608),rgb = (  0.98823529,   0.98823529,   0.99485294),rgb = (  0.98431373,   0.98431373,   0.99338235),rgb = (  0.98039216,   0.98039216,   0.99203431),rgb = (  0.97647059,   0.97647059,   0.99080882),rgb = (  0.97254902,   0.97254902,   0.98970588),rgb = (  0.96862745,   0.96862745,   0.98872549),rgb = (  0.96470588,   0.96470588,   0.98786765),rgb = (  0.96078431,   0.96078431,   0.98713235),rgb = (  0.95686275,   0.95686275,   0.98651961),rgb = (  0.95294118,   0.95294118,   0.98602941),rgb = (  0.94901961,   0.94901961,   0.98566176),rgb = (  0.94509804,   0.94509804,   0.98541667),rgb = (  0.94117647,   0.94117647,   0.98529412),rgb = (  0.93725490,   0.93725490,   0.98529412),rgb = (  0.93333333,   0.93333333,   0.98541667),rgb = (  0.92941176,   0.92941176,   0.98566176),rgb = (  0.92549020,   0.92549020,   0.98602941),rgb = (  0.92156863,   0.92156863,   0.98651961),rgb = (  0.91764706,   0.91764706,   0.98713235),rgb = (  0.91372549,   0.91372549,   0.98786765),rgb = (  0.90980392,   0.90980392,   0.98872549),rgb = (  0.90588235,   0.90588235,   0.98970588),rgb = (  0.90196078,   0.90196078,   0.99080882),rgb = (  0.89803922,   0.89803922,   0.99203431),rgb = (  0.89411765,   0.89411765,   0.99338235),rgb = (  0.89019608,   0.89019608,   0.99485294),rgb = (  0.88627451,   0.88627451,   0.99644608),rgb = (  0.88235294,   0.88235294,   0.99816176),rgb = (  0.87843137,   0.87843137,   1.00000000),rgb = (  0.87450980,   0.87647059,   1.00000000),rgb = (  0.87058824,   0.87463235,   1.00000000),rgb = (  0.86666667,   0.87291667,   1.00000000),rgb = (  0.86274510,   0.87132353,   1.00000000),rgb = (  0.85882353,   0.86985294,   1.00000000),rgb = (  0.85490196,   0.86850490,   1.00000000),rgb = (  0.85098039,   0.86727941,   1.00000000),rgb = (  0.84705882,   0.86617647,   1.00000000),rgb = (  0.84313725,   0.86519608,   1.00000000),rgb = (  0.83921569,   0.86433824,   1.00000000),rgb = (  0.83529412,   0.86360294,   1.00000000),rgb = (  0.83137255,   0.86299020,   1.00000000),rgb = (  0.82745098,   0.86250000,   1.00000000),rgb = (  0.82352941,   0.86213235,   1.00000000),rgb = (  0.81960784,   0.86188725,   1.00000000),rgb = (  0.81568627,   0.86176471,   1.00000000),rgb = (  0.81176471,   0.86176471,   1.00000000),rgb = (  0.80784314,   0.86188725,   1.00000000),rgb = (  0.80392157,   0.86213235,   1.00000000),rgb = (  0.80000000,   0.86250000,   1.00000000),rgb = (  0.79607843,   0.86299020,   1.00000000),rgb = (  0.79215686,   0.86360294,   1.00000000),rgb = (  0.78823529,   0.86433824,   1.00000000),rgb = (  0.78431373,   0.86519608,   1.00000000),rgb = (  0.78039216,   0.86617647,   1.00000000),rgb = (  0.77647059,   0.86727941,   1.00000000),rgb = (  0.77254902,   0.86850490,   1.00000000),rgb = (  0.76862745,   0.86985294,   1.00000000),rgb = (  0.76470588,   0.87132353,   1.00000000),rgb = (  0.76078431,   0.87291667,   1.00000000),rgb = (  0.75686275,   0.87463235,   1.00000000),rgb = (  0.75294118,   0.87647059,   1.00000000),rgb = (  0.74901961,   0.87843137,   1.00000000),rgb = (  0.74509804,   0.88051471,   1.00000000),rgb = (  0.74117647,   0.88272059,   1.00000000),rgb = (  0.73725490,   0.88504902,   1.00000000),rgb = (  0.73333333,   0.88750000,   1.00000000),rgb = (  0.72941176,   0.89007353,   1.00000000),rgb = (  0.72549020,   0.89276961,   1.00000000),rgb = (  0.72156863,   0.89558824,   1.00000000),rgb = (  0.71764706,   0.89852941,   1.00000000),rgb = (  0.71372549,   0.90159314,   1.00000000),rgb = (  0.70980392,   0.90477941,   1.00000000),rgb = (  0.70588235,   0.90808824,   1.00000000),rgb = (  0.70196078,   0.91151961,   1.00000000),rgb = (  0.69803922,   0.91507353,   1.00000000),rgb = (  0.69411765,   0.91875000,   1.00000000),rgb = (  0.69019608,   0.92254902,   1.00000000),rgb = (  0.68627451,   0.92647059,   1.00000000),rgb = (  0.68235294,   0.93051471,   1.00000000),rgb = (  0.67843137,   0.93468137,   1.00000000),rgb = (  0.67450980,   0.93897059,   1.00000000),rgb = (  0.67058824,   0.94338235,   1.00000000),rgb = (  0.66666667,   0.94791667,   1.00000000),rgb = (  0.66274510,   0.95257353,   1.00000000),rgb = (  0.65882353,   0.95735294,   1.00000000),rgb = (  0.65490196,   0.96225490,   1.00000000),rgb = (  0.65098039,   0.96727941,   1.00000000),rgb = (  0.64705882,   0.97242647,   1.00000000),rgb = (  0.64313725,   0.97769608,   1.00000000),rgb = (  0.63921569,   0.98308824,   1.00000000),rgb = (  0.63529412,   0.98860294,   1.00000000),rgb = (  0.63137255,   0.99424020,   1.00000000),rgb = (  0.62745098,   1.00000000,   1.00000000),rgb = (  0.62941176,   1.00000000,   0.99411765),rgb = (  0.63149510,   1.00000000,   0.98811275),rgb = (  0.63370098,   1.00000000,   0.98198529),rgb = (  0.63602941,   1.00000000,   0.97573529),rgb = (  0.63848039,   1.00000000,   0.96936275),rgb = (  0.64105392,   1.00000000,   0.96286765),rgb = (  0.64375000,   1.00000000,   0.95625000),rgb = (  0.64656863,   1.00000000,   0.94950980),rgb = (  0.64950980,   1.00000000,   0.94264706),rgb = (  0.65257353,   1.00000000,   0.93566176),rgb = (  0.65575980,   1.00000000,   0.92855392),rgb = (  0.65906863,   1.00000000,   0.92132353),rgb = (  0.66250000,   1.00000000,   0.91397059),rgb = (  0.66605392,   1.00000000,   0.90649510),rgb = (  0.66973039,   1.00000000,   0.89889706),rgb = (  0.67352941,   1.00000000,   0.89117647),rgb = (  0.67745098,   1.00000000,   0.88333333),rgb = (  0.68149510,   1.00000000,   0.87536765),rgb = (  0.68566176,   1.00000000,   0.86727941),rgb = (  0.68995098,   1.00000000,   0.85906863),rgb = (  0.69436275,   1.00000000,   0.85073529),rgb = (  0.69889706,   1.00000000,   0.84227941),rgb = (  0.70355392,   1.00000000,   0.83370098),rgb = (  0.70833333,   1.00000000,   0.82500000),rgb = (  0.71323529,   1.00000000,   0.81617647),rgb = (  0.71825980,   1.00000000,   0.80723039),rgb = (  0.72340686,   1.00000000,   0.79816176),rgb = (  0.72867647,   1.00000000,   0.78897059),rgb = (  0.73406863,   1.00000000,   0.77965686),rgb = (  0.73958333,   1.00000000,   0.77022059),rgb = (  0.74522059,   1.00000000,   0.76066176),rgb = (  0.75098039,   1.00000000,   0.75098039),rgb = (  0.75686275,   1.00000000,   0.74117647),rgb = (  0.76286765,   1.00000000,   0.73125000),rgb = (  0.76899510,   1.00000000,   0.72120098),rgb = (  0.77524510,   1.00000000,   0.71102941),rgb = (  0.78161765,   1.00000000,   0.70073529),rgb = (  0.78811275,   1.00000000,   0.69031863),rgb = (  0.79473039,   1.00000000,   0.67977941),rgb = (  0.80147059,   1.00000000,   0.66911765),rgb = (  0.80833333,   1.00000000,   0.65833333),rgb = (  0.81531863,   1.00000000,   0.64742647),rgb = (  0.82242647,   1.00000000,   0.63639706),rgb = (  0.82965686,   1.00000000,   0.62524510),rgb = (  0.83700980,   1.00000000,   0.61397059),rgb = (  0.84448529,   1.00000000,   0.60257353),rgb = (  0.85208333,   1.00000000,   0.59105392),rgb = (  0.85980392,   1.00000000,   0.57941176),rgb = (  0.86764706,   1.00000000,   0.56764706),rgb = (  0.87561275,   1.00000000,   0.55575980),rgb = (  0.88370098,   1.00000000,   0.54375000),rgb = (  0.89191176,   1.00000000,   0.53161765),rgb = (  0.90024510,   1.00000000,   0.51936275),rgb = (  0.90870098,   1.00000000,   0.50698529),rgb = (  0.91727941,   1.00000000,   0.49448529),rgb = (  0.92598039,   1.00000000,   0.48186275),rgb = (  0.93480392,   1.00000000,   0.46911765),rgb = (  0.94375000,   1.00000000,   0.45625000),rgb = (  0.95281863,   1.00000000,   0.44325980),rgb = (  0.96200980,   1.00000000,   0.43014706),rgb = (  0.97132353,   1.00000000,   0.41691176),rgb = (  0.98075980,   1.00000000,   0.40355392),rgb = (  0.99031863,   1.00000000,   0.39007353),rgb = (  1.00000000,   1.00000000,   0.37647059),rgb = (  1.00000000,   0.99019608,   0.37254902),rgb = (  1.00000000,   0.98026961,   0.36862745),rgb = (  1.00000000,   0.97022059,   0.36470588),rgb = (  1.00000000,   0.96004902,   0.36078431),rgb = (  1.00000000,   0.94975490,   0.35686275),rgb = (  1.00000000,   0.93933824,   0.35294118),rgb = (  1.00000000,   0.92879902,   0.34901961),rgb = (  1.00000000,   0.91813725,   0.34509804),rgb = (  1.00000000,   0.90735294,   0.34117647),rgb = (  1.00000000,   0.89644608,   0.33725490),rgb = (  1.00000000,   0.88541667,   0.33333333),rgb = (  1.00000000,   0.87426471,   0.32941176),rgb = (  1.00000000,   0.86299020,   0.32549020),rgb = (  1.00000000,   0.85159314,   0.32156863),rgb = (  1.00000000,   0.84007353,   0.31764706),rgb = (  1.00000000,   0.82843137,   0.31372549),rgb = (  1.00000000,   0.81666667,   0.30980392),rgb = (  1.00000000,   0.80477941,   0.30588235),rgb = (  1.00000000,   0.79276961,   0.30196078),rgb = (  1.00000000,   0.78063725,   0.29803922),rgb = (  1.00000000,   0.76838235,   0.29411765),rgb = (  1.00000000,   0.75600490,   0.29019608),rgb = (  1.00000000,   0.74350490,   0.28627451),rgb = (  1.00000000,   0.73088235,   0.28235294),rgb = (  1.00000000,   0.71813725,   0.27843137),rgb = (  1.00000000,   0.70526961,   0.27450980),rgb = (  1.00000000,   0.69227941,   0.27058824),rgb = (  1.00000000,   0.67916667,   0.26666667),rgb = (  1.00000000,   0.66593137,   0.26274510),rgb = (  1.00000000,   0.65257353,   0.25882353),rgb = (  1.00000000,   0.63909314,   0.25490196),rgb = (  1.00000000,   0.62549020,   0.25098039),rgb = (  1.00000000,   0.61176471,   0.24705882),rgb = (  1.00000000,   0.59791667,   0.24313725),rgb = (  1.00000000,   0.58394608,   0.23921569),rgb = (  1.00000000,   0.56985294,   0.23529412),rgb = (  1.00000000,   0.55563725,   0.23137255),rgb = (  1.00000000,   0.54129902,   0.22745098),rgb = (  1.00000000,   0.52683824,   0.22352941),rgb = (  1.00000000,   0.51225490,   0.21960784),rgb = (  1.00000000,   0.49754902,   0.21568627),rgb = (  1.00000000,   0.48272059,   0.21176471),rgb = (  1.00000000,   0.46776961,   0.20784314),rgb = (  1.00000000,   0.45269608,   0.20392157),rgb = (  1.00000000,   0.43750000,   0.20000000),rgb = (  1.00000000,   0.42218137,   0.19607843),rgb = (  1.00000000,   0.40674020,   0.19215686),rgb = (  1.00000000,   0.39117647,   0.18823529),rgb = (  1.00000000,   0.37549020,   0.18431373),rgb = (  1.00000000,   0.35968137,   0.18039216),rgb = (  1.00000000,   0.34375000,   0.17647059),rgb = (  1.00000000,   0.32769608,   0.17254902),rgb = (  1.00000000,   0.31151961,   0.16862745),rgb = (  1.00000000,   0.29522059,   0.16470588),rgb = (  1.00000000,   0.27879902,   0.16078431),rgb = (  1.00000000,   0.26225490,   0.15686275),rgb = (  1.00000000,   0.24558824,   0.15294118),rgb = (  1.00000000,   0.22879902,   0.14901961),rgb = (  1.00000000,   0.21188725,   0.14509804),rgb = (  1.00000000,   0.19485294,   0.14117647),rgb = (  1.00000000,   0.17769608,   0.13725490),rgb = (  1.00000000,   0.16041667,   0.13333333),rgb = (  1.00000000,   0.14301471,   0.12941176),rgb = (  1.00000000,   0.12549020,   0.12549020),rgb = (  0.98627451,   0.12156863,   0.12156863),rgb = (  0.97242647,   0.11764706,   0.11764706),rgb = (  0.95845588,   0.11372549,   0.11372549),rgb = (  0.94436275,   0.10980392,   0.10980392),rgb = (  0.93014706,   0.10588235,   0.10588235),rgb = (  0.91580882,   0.10196078,   0.10196078),rgb = (  0.90134804,   0.09803922,   0.09803922),rgb = (  0.88676471,   0.09411765,   0.09411765),rgb = (  0.87205882,   0.09019608,   0.09019608),rgb = (  0.85723039,   0.08627451,   0.08627451),rgb = (  0.84227941,   0.08235294,   0.08235294),rgb = (  0.82720588,   0.07843137,   0.07843137),rgb = (  0.81200980,   0.07450980,   0.07450980),rgb = (  0.79669118,   0.07058824,   0.07058824),rgb = (  0.78125000,   0.06666667,   0.06666667),rgb = (  0.76568627,   0.06274510,   0.06274510),rgb = (  0.75000000,   0.05882353,   0.05882353),rgb = (  0.73419118,   0.05490196,   0.05490196),rgb = (  0.71825980,   0.05098039,   0.05098039),rgb = (  0.70220588,   0.04705882,   0.04705882),rgb = (  0.68602941,   0.04313725,   0.04313725),rgb = (  0.66973039,   0.03921569,   0.03921569),rgb = (  0.65330882,   0.03529412,   0.03529412),rgb = (  0.63676471,   0.03137255,   0.03137255),rgb = (  0.62009804,   0.02745098,   0.02745098),rgb = (  0.60330882,   0.02352941,   0.02352941),rgb = (  0.58639706,   0.01960784,   0.01960784),rgb = (  0.56936275,   0.01568627,   0.01568627),rgb = (  0.55220588,   0.01176471,   0.01176471),rgb = (  0.53492647,   0.00784314,   0.00784314),rgb = (  0.51752451,   0.00392157,   0.00392157),rgb = (  0.50000000,   0.00000000,   0.00000000)}
}

\definecolor{mittelblau}{RGB}{0, 126, 198}
\definecolor{violettblau}{cmyk}{0.9, 0.6, 0, 0}
\definecolor{rot}{RGB}{238, 28 35}
\definecolor{apfelgruen}{RGB}{140, 198, 62}
\definecolor{gelb}{RGB}{1, 221, 0}
\definecolor{orange}{RGB}{244, 111, 33}
\definecolor{pink}{RGB}{237, 0, 140}
\definecolor{lila}{RGB}{128, 10, 145}
\definecolor{hellgrau}{RGB}{224, 224, 224}
\definecolor{mittelgrau}{RGB}{128, 128, 128}
\definecolor{dunkelgrau}{RGB}{80,80,80}
\definecolor{anthrazit}{RGB}{19, 31, 31}

\definecolor{myblue}{RGB}{80,80,160} 
\definecolor{mygreen}{RGB}{80,160,80}
\definecolor{myorgange}{RGB}{204,102,0}
\definecolor{lightblue}{RGB}{51,153,255}

\addto\captionsenglish{}

\makeatother

\begin{document}

\title{Near-Capacity Detection and Decoding: Code Design for Dynamic User
Loads in Gaussian Multiple Access Channels }

\author{Xiaojie~Wang,~\IEEEmembership{Student~Member,~IEEE,} Sebastian~Cammerer,~\IEEEmembership{Student~Member,~IEEE,}
and~Stephan~ten~Brink,~\IEEEmembership{Senior~Member,~IEEE}\thanks{Parts of the results were presented at the International Symposium on Turbo Codes and Iterative Information Processing (ISTC), Hong Kong, December 2018.}\thanks{The authors are with Institute of Telecommunications, Pfaffenwaldring 47, University of Stuttgart, 70569 Stuttgart, Germany (e-mail: \{wang, cammerer, tenbrink\}@inue.uni-stuttgart.de).}}
\maketitle
\begin{abstract}
This paper considers the forward error correction (FEC) code design
for approaching the capacity of a \emph{dynamic} multiple access channel
(MAC) where both the number of users and their respective signal powers
keep constantly changing, resembling the scenario of an actual wireless
cellular system. To obtain a low-complexity non-orthogonal multiple
access (NOMA) scheme, we propose a serial concatenation of a low-density
parity-check (LDPC) code and a repetition code (REP), this way achieving
near Gaussian MAC (GMAC) capacity performance while coping with the
dynamics of the MAC system. The joint optimization of the LDPC and
REP codes is addressed by matching the analytical extrinsic information
transfer (EXIT) functions of the sub-optimal multi-user detector (MUD)
and the channel code for a specific and static MAC system, achieving
near-GMAC capacity. We show that the near-capacity performance can
be flexibly maintained with the same LDPC code regardless of the variations
in the number of users and power levels. This flexibility (or elasticity)
is provided by the REP code, acting as ``user-load and power equalizer'',
dramatically simplifying the practical implementation of NOMA schemes,
as only a single LDPC code is needed to cope with the dynamics of
the MAC system.
\end{abstract}

\begin{IEEEkeywords}
Interleave division multiple access, EXIT chart, non-orthogonal multiple
access, LDPC code, multi-user detection.
\end{IEEEkeywords}

\section{Introduction}

The fundamental capacity limit of information transmission between
one transmitter and one receiver over an additive white Gaussian noise
(AWGN) channel, laid down by Shannon, has virtually been achieved.
In fact, there exist several powerful and practical codes, e.g., Turbo
\cite{BerrouTurboTC96}, LDPC \cite{RichardsonLDPCDesignTIT01} and
Polar codes \cite{ArikanPolarTIT09}, that can operate close to capacity
with both moderate codeword length as well as feasible decoding complexity.
However, the situation changes when considering multi-user setups,
such as the multiple access channel (MAC), i.e., multiple transmitters
(or users) and one single receiver.

The MAC capacity (region) has been known since the 1970s \cite{CoverITBook06}
and can be determined by a tuple of rates $R_{k},$ $1\leq k\leq N$
of the $N$ individual users. In this paper, we assume that no cooperation
is allowed among users, e.g., no power control and/or cooperative
encoding at the transmitters. We refer interested readers to \cite{cooperativeMACcap,DTseCoopComm,VPoorNOMAcoop}
for a detailed discussion of capacity regions and different cooperative
schemes in multi-user settings. Often, the capacity region is \textit{dominantly
}bounded by the sum capacity constraint of all users. These rate tuples,
resulting in maximum sum-rate, constitute an \textit{edge} or \textit{facet
}in vector space\textit{ }with dimension of two or higher, respectively.
It is referred to as \textit{dominant face }of the MAC capacity region,
which is of particular practical interest. 

\begin{figure}[tbh]
\begin{centering}
\includegraphics[width=0.8\columnwidth]{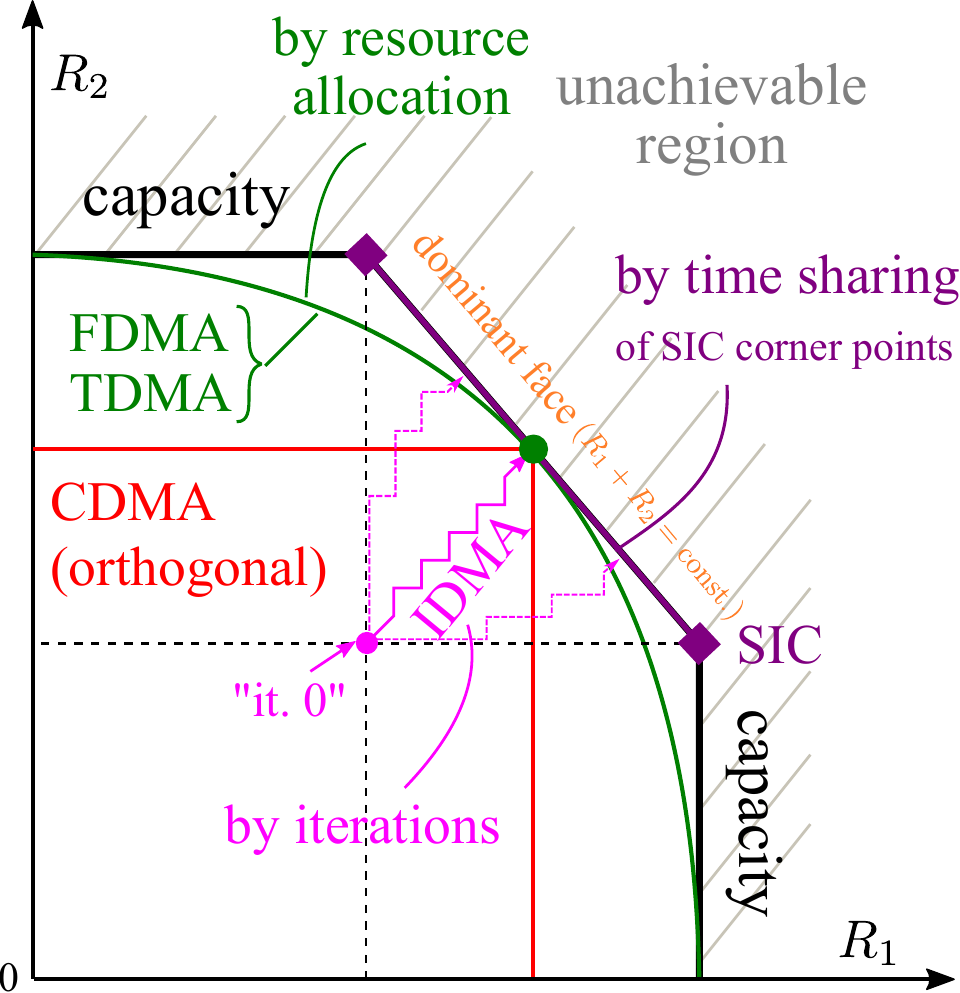}
\par\end{centering}
\caption{Example of two-user Gaussian MAC capacity region with achievable rate
pairs $\left(R_{1},R_{2}\right)$ of various multiple access schemes
with equal transmit power $P_{1}=P_{2}$; the capacity region is denoted
by the black curve.\label{fig:MAC_Cap_Reg}}
\end{figure}

To achieve arbitrary points of the \textit{dominant face} (illustrated
for the two-user case in Fig.~\ref{fig:MAC_Cap_Reg}), existing approaches
rely on joint multi-user decoding or successive interference cancellation
(SIC) with time sharing/rate-splitting \cite{DTseBookFund,KramerNOW,HouICtimingOffset}.
Joint decoding is not feasible in terms of complexity even for a small
number of users. While SIC can achieve certain corner points of the
capacity region, it suffers from high latency and potential error
propagation. Moreover if finite block length codes are used, the gap
to capacity increases as the number of users grows \cite[Fig. 5]{YHuLPNOMAbook}.
However, due to its theoretically capacity-achieving capability with
codeword length approaching infinity, various non-orthogonal multiple
access (NOMA) schemes in the literature still apply SIC to delay-insensitive
applications \cite{NOMA5GCommMag,NOMADingCommMag,NOMAVTC13,NOMADingSPL}. 

Other orthogonal methods such as time division multiple access (TDMA),
frequency division multiple access (FDMA) and code division multiple
access (CDMA) avoid interference by providing orthogonal channels
so that each user can detect and decode its signal independently.
However, these schemes may require, on the one hand, complex synchronization
and tracking algorithms to maintain orthogonality in time/frequency/code
domain, while, on the other hand, achieve solely one point of the
capacity region where optimum resource allocation in terms of bandwidth,
power and time is performed \cite{DTseBookFund,DTseMAC} (see green
dot in Fig.~\ref{fig:MAC_Cap_Reg}).

In \cite{NOMA5GCommMag,NOMASurvey}, the authors study the advantage
of NOMA over orthogonal methods in the context of 5G for supporting
massive connectivity and compare various NOMA schemes with respect
to spectral efficiency, latency and cell-edge throughput. As a special
form of superposition coding, interleave-division multiple-access
(IDMA) \cite{LiPingIDMACL04,LipingIDMATWC06} appears to be a particularly
interesting NOMA scheme that can approach the GMAC capacity, yet with
a simple multi-user detector (MUD) \cite{IDMAPeterHoeher,LLiuICC16,LLiuTSP18}.
It uses interleaved codes and a low-complexity graph-based message
passing algorithm (MPA) and exhibits desired properties such as low
complexity, parallelizable computation and asynchronous transmission.

\subsection{Problem formulation}

 In NOMA, the multiple access interference (MAI) rather than the
noise is the bottleneck. For example, when treating MAI as noise,
only the rate tuple marked with ``it. 0'' in Fig.~\ref{fig:MAC_Cap_Reg}
can be achieved, which is, from an information theoretic perspective,
sub-optimal \cite{CoverITBook06}.

SIC has been known for a long time to achieve the corner points (marked
as diamonds in Fig. \ref{fig:MAC_Cap_Reg}) of the MAC capacity region
\cite{CoverITBook06}. Yet, it requires very accurate channel knowledge
and capacity-achieving code design for every user, each with a different
code rate, i.e., requiring a rate-specific code design. This hinders
its practical deployment since strict user-load, decoding order and
power control are needed. 

It was reported in \cite{LiPingIDMACL04,LipingIDMATWC06} that IDMA
is an attractive method for approaching the MAC capacity with low-rate
codes and low complexity receivers. In IDMA, the same error correction
code may apply to all users. A low complexity parallel interference
cancellation (PIC)-based multi-user detector (MUD) is serially concatenated
(as an inner code component) with the outer channel code in a turbo-like
fashion \cite{774855}. Then, (turbo) iterations between the two components
are carried out to improve the achievable rate toward the MAC capacity.

For close-to-optimal performance of the low complexity PIC-based MUD,
Gaussian-distributed interference is required, an assumption that
holds for a large number of users when a finite alphabet, e.g., binary
phase shift keying (BPSK), is used. In this case, the single user
decoder is seriously challenged by MAI at the beginning of the turbo
iterations due to the large number of users, leading to a very low
initial effective SNR. For this reason, low-rate codes appear to be
inevitable in IDMA systems.

As LDPC codes are capacity-approaching \cite{RichardsonLDPCDesignTIT01}
and there exist several useful tools for their optimization, also
taking into account a detector front-end \cite{tenBrinkLDPCEXITTC04}
in the context of point-to-point (P2P) communications, we use LDPC
codes for channel coding in the IDMA framework. However, such a low-rate
code design and its decoding in the range of, e.g., code rates $R_{c}\leq\frac{1}{10}$
becomes a very challenging task \cite{AndriyanovaLowRateCodeTC12}.
The reasons are twofold: firstly, the decoding graph becomes relatively
dense at low-rates; secondly, the Gaussian approximation (GA) of messages
at the check nodes (CN) becomes inaccurate due to the necessity of
low-degree nodes \cite{AndriyanovaLowRateCodeTC12}. Also, the maximum
node degrees should not exceed a certain limit to maintain feasible
finite length code constructions. This renders code design for IDMA
systems to be a challenging and interesting research direction.

The existing approaches for designing capacity-achieving codes in
MAC systems require precise knowledge about the number of users and
the received signal power. Typically, both the number of users and
their respectively received signal power can vary significantly (``dynamic''
MAC system) while an adaptive code optimization is not desirable for
the hardware implementation of the decoder. The hardware structure
is often optimized and can not be easily re-programmed on the fly.
Therefore, an intriguing question is how to maintain near-capacity
performance under a \textit{dynamic} MAC system based on the \textit{same}
channel code design at each user.

\subsection{Related work}

Previous works \cite{LiPingIDMACL04,LipingIDMATWC06,BoutrosCaireTIT02,CaireTIT04}
show that iterative detection and decoding can achieve close to MAC
capacity. In \cite{KLiEXITTurboMUTWC05,KLiIDMAOptTWC07,JSACCDMAChipInt}
extrinsic information transfer (EXIT) charts \cite{tenBrinkEXITTC01}
are introduced to analyze the performance of iterative MUD and channel
decoding; therein, the main focus is on different multi-user detectors.
A power allocation scheme among users is proposed to close the gap
to the MAC capacity. References \cite{RZhangCodeSpreadingIDMAWCNC07,ZhangIDMAHybTVT10}
consider the coding-spreading trade-off and a hybrid MUD to improve
convergence, yet only based on a few selected codes and a few number
of users. However, the systematic design and optimization of the joint
detection and decoding schemes over the multiple-access channel still
remains an open problem.

Recent works in \cite{ABSTWCL182userLDPC,DuCLLDPC2user,SharifiTWC1662userLDPC}
address LDPC code design with two users applying optimal joint detection
and a Gaussian-mixture model to approximate the message probability
density function (PDF) at the CNs. In \cite{ZhangIDMALDPCOptGCwks17},
the LDPC codes are optimized for IDMA systems with a relatively large
number of users (6 to 10), again with an optimum joint MUD. In \cite{ZhangYSIDMA5GCodeCL18},
the 5G LDPC code is studied for multi-user communications assuming
optimal MUD. This joint MUD eliminates one of the most attractive
features of IDMA, namely, its low complexity. Furthermore, for a moderate
to large number of users, an optimum joint MUD is computationally
prohibitive and, therefore, the code design for such large systems
has not yet been adequately addressed, although the design of binary
LDPC codes for iterative detection and decoding is well-understood
in the EXIT-chart framework \cite{tenBrinkLDPCEXITTC04,SchmalenBrinkSCC13,CammererISTC16}.

In \cite{GSongIT16,GSongISIT15}, the authors study the code properties
in the MAC setting and optimize irregular repeat and accumulate (IRA)
code design for the sub-optimal MUD of IDMA a with fixed number of
users and an equal power setting. In \cite{HuYangCLIDMAlowcost,ChulongCLIDMASCMA},
low-cost implementations of IDMA are considered and compared to other
NOMA schemes, showing the superior performance of IDMA.

\subsection{Contributions}

This paper considers a dynamic (uplink) IDMA system with both varying
number of users and varying received signal power strength. We seek
to find a \textit{simple} and \textit{flexible} coding scheme for
coping with the dynamics of the system parameters. The contributions
of this paper are summarized in the following.
\begin{itemize}
\item We propose to use serially concatenated LDPC and repetition (REP)
codes for a dynamic MAC system. We refer to a dynamic MAC system as
a model for a real cellular network with constantly changing number
of users and varying signal power levels.
\item An analytical expression of the EXIT function capturing the LDPC,
REP and the sub-optimal MUD is derived.
\item A joint optimization of LDPC and REP code is carried out to match
the low-complexity MUD in a given (static) MAC system.
\item We show that the REP code can be used as ``user-load equalizer''
to compensate the influence of a varying number of users, i.e., the
\textit{same} LDPC code fits different number of active users with
hardly any performance loss. 
\item We show that the REP code can also be used as ``power equalizer''
to compensate the influence of varying power levels, i.e., the \textit{same}
LDPC code fits different power levels with hardly any performance
loss.
\end{itemize}
\begin{figure*}[tp]
\begin{centering}
\resizebox {1.96\columnwidth} {!} {

\begin{tikzpicture}[scale=0.096,ampersand replacement=\&]
\matrix (TXchains) [row sep=2.5mm, column sep=4mm] { 		
\node[]  () {};=\&
\node[]  () {LDPC Encoder};\&
\node[text width=4cm, align = center]  () {REP Encoder};\&
\node[]  () {Interleaver};\&
\node[align=center]  () {Mapper \\ and \\ Phase-Scrambler};\&
\node[]  () {Channel};\&
\&
\\
\node[]  (bitsource1) {$\mathbf{b}_{1}$};\&
\node[dspsquare,very thick,draw = myorgange]       (fec1) {$R_{\mathrm{c}}$}; \& 		
\node[dspsquare,very thick,draw = mygreen]       (ce1) {$R_{\mathrm{r}}$}; \& 		
\node[dspsquare,thick]       (interleaver1) {\,$\boldsymbol{\pi}_{1}$\,}; \&		
\node[dspsquare,very thick,draw = lightblue]   (mapper1) {$\boldsymbol{\varphi}_{1}$}; \&
\node[dspsquare,thick]   (channel1) {}; \&
\&
\\
\node[]  () {};\&				
\node[]  () {$\vdots$};\&
\node[]  () {$\vdots$};\&
\node[]  () {$\vdots$};\&
\node[]  () {$\vdots$};\&
\node[]  () {$\vdots$};\&
\node[dspadder]                     (adder) {};\&
\\
\node[] ()  (bitsourceN) {$\mathbf{b}_{N}$};\& 
\node[dspsquare,very thick,draw = myorgange]       (fecN) {}; \& 		
\node[dspsquare,very thick,draw = mygreen]       (ceN) {}; \& 		
\node[dspsquare,thick]       (interleaverN) {\,$\boldsymbol{\pi}_{N}$\,}; \&		
\node[dspsquare,very thick,draw = lightblue]   (mapperN) {$\boldsymbol{\varphi}_{N}$}; \&
\node[dspsquare,thick]   (channelN) {}; \&
\&				
\\
\node[]  () {};\&
\node[] () {};\&
\node[] () {};\&
\node[]  () {};\&
\node[]   () {};\&
\node[]  () {};\&
\node[]   () {};\&
\\
\node[]  () {};\&
\node[]  () {LDPC Decoder};\&
\node[]   () {REP Decoder};\&
\node[]  () {Deinterleaver};\&
\node[align=center]   () {Demapper \\ and \\ De-scrambler};\&
\node[]   () {};\&
\node[]   () {};\&
\\
\node[] ()  (bitsinke1) {$\hat{\mathbf{b}}_{1}$};\&
\node[dspsquare,very thick,draw = myorgange]   (Noisedecoder1) {$R_{\mathrm{c}}$}; \&
\node[dspsquare,very thick,draw = mygreen]   (decoder1) {$R_{\mathrm{r}}$}; \&
\node[dspsquare,thick]       (deinterleaver1) {\,$\boldsymbol{\pi}_{1}^{-1}$\,}; \&
\node[dspsquare,very thick,draw = lightblue]       (demapper1) {$\boldsymbol{\varphi}^{-1}_{1}$}; \&			
\\
\node[]  () {};\&
\node[]  () {$\vdots$};\&
\node[]  () {$\vdots$};\&
\node[]  () {$\vdots$};\&
\node[]  (cNode) {$\vdots$};\&
\&
\\	
\node[]  (bitsinkeN) {$\hat{\mathbf{b}}_{N}$};\&
\node[dspsquare,very thick,draw = myorgange]   (NoisedecoderN) {}; \&	
\node[dspsquare,very thick,draw = mygreen]   (decoderN) {}; \&
\node[dspsquare,thick]       (deinterleaverN) {\,$\boldsymbol{\pi}_{N}^{-1}$\,}; \&
\node[dspsquare,very thick,draw = lightblue]       (demapperN) {$\boldsymbol{\varphi}^{-1}_{N}$}; \& 			
\\
};		
\node[dspfilter,very thick,right=of cNode,xshift=5mm,yshift=0mm,minimum height=35mm,draw = myblue]                     (MUD) {SoIC- \\ MUD};

		
	\draw[dspconn] (bitsource1) -- 				(fec1);
    \draw[dspconn] (fec1) -- 				      (ce1);
	\draw[dspconn] (ce1) -- 				(interleaver1);
    \draw[dspconn] (interleaver1) -- 				(mapper1);
    \draw[dspconn] (mapper1) -- 				(channel1) node[midway,above] {$\tilde{\mathbf{x}}_{1}$};
    \draw[dspconn] (channel1.east) -- 				(adder);
	\draw[dspconn] (bitsourceN) -- 				(fecN);
    \draw[dspconn] (fecN) -- 				      (ceN);
	\draw[dspconn] (ceN) -- 				(interleaverN);
    \draw[dspconn] (interleaverN) -- 				(mapperN);
    \draw[dspconn] (mapperN) -- 				(channelN) node[midway,above] {$\tilde{\mathbf{x}}_{n}$};
    \draw[dspconn] (channelN.east) -- 				(adder);
    \draw[dspconn,<-] (adder)  -- +(0,8) node[near end,above, yshift=2mm] {AWGN $\boldsymbol{n}$};
     \draw[dspconn] (adder.east) |- +(1,0) |- (MUD.east) node[near end,right, xshift=-1mm,yshift=2mm] {$\mathbf{y}$};
    \draw[dspconn] ($(MUD.west)+(0,12.5)$) -- node[above] {$\mathbf{y}_{1}$} 		+(-12.5,0);
    \draw[dspconn,<-] ($(MUD.west)+(0,9.5)$) -- node[below] {$\hat{\mathbf{x}}_{1}$}		+(-12.5,0);
    \draw[dspconn] ($(demapper1.west)+(0,1.5)$) -- 				($(deinterleaver1.east)+(0,1.5)$) node[midway,above] {$\mathbf{L}^{\mathrm{E}}_{\mathrm{M},1}$};
    \draw[dspconn] ($(deinterleaver1.east)+(0,-1.5)$) -- 				($(demapper1.west)+(0,-1.5)$) node[midway,below] {$\mathbf{L}^{\mathrm{A}}_{\mathrm{M},1}$};
    \draw[dspconn] ($(deinterleaver1.west)+(0,1.5)$) -- 				($(decoder1.east)+(0,1.5)$) node[midway,above] {$\mathbf{L}^{\mathrm{A}}_{\mathrm{R\leftarrow M},1}=\tilde{\mathbf{L}}^{\mathrm{E}}_{\mathrm{M},1}$};
    \draw[dspconn] ($(decoder1.east)+(0,-1.5)$) -- 				($(deinterleaver1.west)+(0,-1.5)$) node[midway,below] {$\mathbf{L}^{\mathrm{E}}_{\mathrm{R\rightarrow M},1}=\tilde{\mathbf{L}}^{\mathrm{A}}_{\mathrm{M},1}$};
    \draw[dspconn] ($(decoder1.west)+(0,1.5)$) -- 				($(Noisedecoder1.east)+(0,1.5)$) node[midway,above] {$\mathbf{L}^{\mathrm{A}}_{\mathrm{D},1}=\mathbf{L}^{\mathrm{E}}_{\mathrm{R\rightarrow D},1}$};
    \draw[dspconn] ($(Noisedecoder1.east)+(0,-1.5)$) -- 				($(decoder1.west)+(0,-1.5)$) node[midway,below] {$\mathbf{L}^{\mathrm{E}}_{\mathrm{D},1}=\mathbf{L}^{\mathrm{A}}_{\mathrm{R\leftarrow D},1}$};
	\draw[dspconn] (Noisedecoder1) -- 				(bitsinke1);
    \draw[dspconn,<-] ($(MUD.west)+(0,-12.5)$) -- 		+(-12.5,0);
    \draw[dspconn] ($(MUD.west)+(0,-9.5)$) -- 		+(-12.5,0);
    \draw[dspconn] ($(demapperN.west)+(0,1.5)$) -- 				($(deinterleaverN.east)+(0,1.5)$);
    \draw[dspconn] ($(deinterleaverN.east)+(0,-1.5)$) -- 				($(demapperN.west)+(0,-1.5)$);
    \draw[dspconn] ($(deinterleaverN.west)+(0,1.5)$) -- 				($(decoderN.east)+(0,1.5)$);
    \draw[dspconn] ($(decoderN.east)+(0,-1.5)$) -- 				($(deinterleaverN.west)+(0,-1.5)$);
    \draw[dspconn] ($(decoderN.west)+(0,1.5)$) -- 				($(NoisedecoderN.east)+(0,1.5)$);
    \draw[dspconn] ($(NoisedecoderN.east)+(0,-1.5)$) -- 				($(decoderN.west)+(0,-1.5)$);
    \draw[dspconn] (NoisedecoderN) -- 				(bitsinkeN);
\end{tikzpicture}
}
\par\end{centering}
\caption{IDMA system model; all users have the same power/code/modulation;
note that boldface letters are used to denote vectors.\label{fig:IDMA-System-Model}}
\end{figure*}
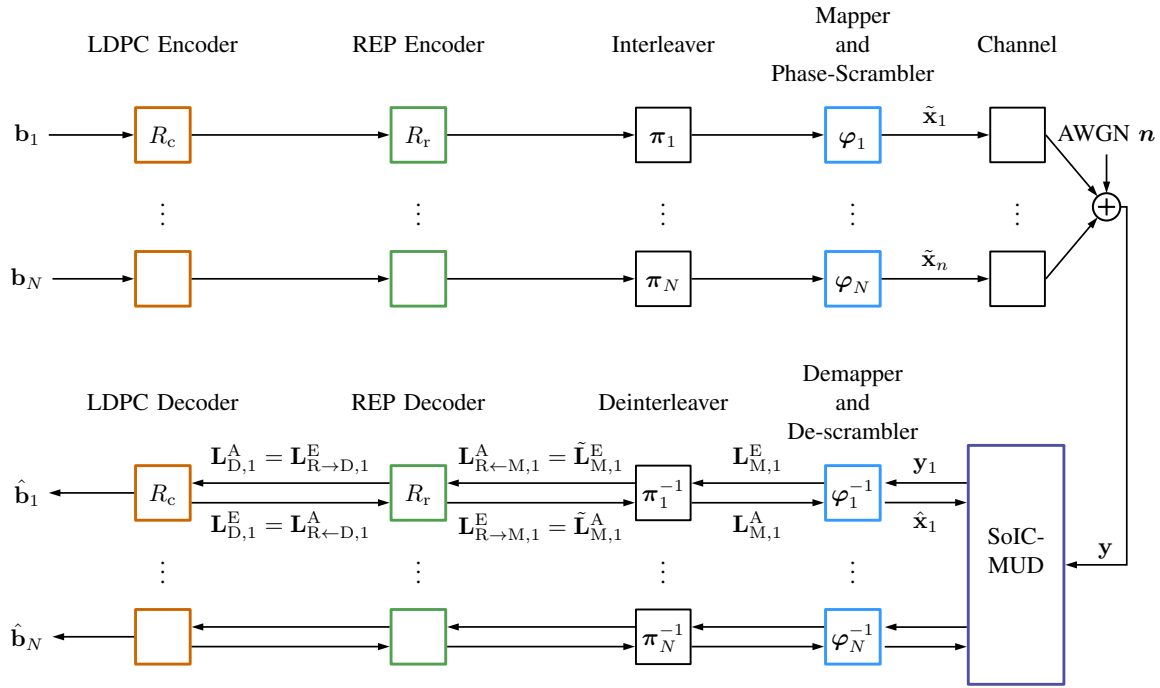

Simulation results show that we can achieve close to GMAC capacity
with $\unit[1.28]{dB}$ $\frac{E_{b}}{N_{0}}$ loss at a sum-rate
of $\unit[0.9375]{bpcu}$ supporting $30$ users each with BPSK modulation\footnote{The target sum-rate is chosen such that the interference and noise
have roughly the same impact on the performance. For low SNRs, the
performance is mainly noise-limited and single-user coding suffices.
For high SNRs, the performance is mainly interference-limited and
orthogonal schemes or repetition coding suffice(s); Detailed discussions
in Sec. \ref{subsec:Operation-Regime}.}; this performance can be achieved over a wide range of number of
users and in equal or non equal-power scenarios by simply varying
the repetition code-rate while fixing the LDPC code.

The paper is structured as follows. The IDMA system model and the
multi-stage low-complexity iterative receiver is explained in Sec.
\ref{sec:IDMA-System-Model}. The EXIT functions are subsequently
reviewed and elaborated on in the IDMA framework in Sec. \ref{sec:EXIT-Analysis}.
Then, the joint optimization of the serially concatenated LDPC and
REP code is presented in Sec. \ref{sec:Degree-Profile-Optimization}.
Sec. \ref{sec:Beyond-Equal-Power} discusses the performance in unequal
power scenarios by exploiting the ``power equalizer'' functionality
of the REP code. Finally, Sec. \ref{sec:Conclusion} concludes this
paper.

\section{\label{sec:IDMA-System-Model}IDMA System Model}

Fig.~\ref{fig:IDMA-System-Model} shows the IDMA system model with
$N$ non-cooperative users. Each user encodes and decodes its data
separately using an LDPC encoder of code rate of $R_{c}$ and a serially
concatenated repetition code of rate of $R_{r}=\frac{1}{d_{\mathrm{r}}}$.
Note that the code parameters, e.g., degree profile and parity check
matrix $\mathbf{H}$ are the same among all the users. The total code
rate is $R_{\mathrm{tot}}=R_{c}R_{r}$. The interleaver is, on the
contrary, user-specific to allow efficient user separation at the
receiver. After interleaving, the coded bits are mapped to symbols,
e.g., BPSK, and transmitted over a channel. 

Consider uncorrelated ergodic Rayleigh fading channels, the $m$th
received signal (i.e., the $m$th element of $\mathbf{y}$ in Fig.~\ref{fig:IDMA-System-Model})
of all users can be written as
\begin{equation}
y_{m}={\displaystyle \sum_{i=1}^{N}}\sqrt{P_{i}}h_{i,m}\underset{:=\widetilde{x}_{i,m}}{\underbrace{x_{i,m}\cdot e^{j\varphi_{i,m}}}}+n_{m}
\end{equation}
where $m$ is the discrete-time index, $n_{m}$ is circularly symmetric
(complex-valued) AWGN with zero mean and variance $\sigma_{n}^{2}$,
$h_{i,m}$ is the uncorrelated (both over time and among different
users) small-scale Rayleigh fading channel coefficient, $P_{i}$ is
the received signal power strength of the $i$th user and $\varphi_{i,m}$
is a pseudo random phase scrambling to avoid ambiguity of the super-constellation
(Cartesian product of all users' constellations). This random phase
shift could also be the consequence of, e.g., the channel and/or explicit
``scrambling'' and we include this into each user's mapper (only
for AWGN channels; for Rayleigh fading channels, this step can be
omitted). Throughout this paper, the phases $\varphi_{i,m}$ are independently
and uniformly distributed in $\left[0,\pi\right)$. Alternatively,
a phase ``scrambling'', which is a constant for all symbols but
distinct per user, can also be applied (but subject to optimization).
The output of the mapper of the $i$th user with BPSK modulation at
the $m$th time instant is $\tilde{x}_{i,m}\in\left\{ \pm e^{j\varphi_{i,m}}\right\} $.
This ``phase scrambling'' can improve the superimposed multi-user
codeword distance \cite{GSongIT16}, particularly in AWGN channels.
For notational brevity, the symbol index $m$ is dropped in the following.

The received signal is first processed by a MUD. An optimum MUD is
to maximize the \textit{a posteriori} probability (APP) of each bit,
i.e., 
\begin{equation}
L_{M,i}\left(\left.x_{i}\right|y_{m}\right)=\log\frac{\underset{\mathbf{s}\in\mathcal{S}_{i,+1}}{\sum}p\left(\left.\mathbf{s}\right|y_{m}\right)}{\underset{\mathbf{s}\in\mathcal{S}_{i,-1}}{\sum}p\left(\left.\mathbf{s}\right|y_{m}\right)}\label{eq:APPDec}
\end{equation}
where $\mathbf{s}=\left[x_{1},x_{2},\cdots,x_{N}\right]$ denotes
a symbol vector of all users, $\mathcal{S}_{i,+1}$ and $\mathcal{S}_{i,-1}$
with the cardinality $2^{N-1}$ are two sets containing the symbol
vectors with $\left[\mathbf{s}\right]_{i}=x_{i}=+1$ and $x_{i}=-1$,
respectively. This requires a complexity of $O\left(M^{N}\right)$
where $M$ denotes the number of constellation symbols per user. The
exponentially increasing complexity with the number of users $N$
prohibits its practical implementation. Therefore, a sub-optimal soft
interference cancellation (SoIC) based low complexity MUD was proposed
in \cite{LiPingIDMACL04}. The sub-optimal MUD first cancels out other
users' signals; for instance with BPSK signaling, the $i$th user's
signal is estimated by the conditional mean estimate \cite{GuoSVMMSEmIITI05}
\begin{equation}
\hat{x}_{i}=\mathrm{E}\left[\left.x_{i}\right|L_{\mathrm{M,}i}^{\mathrm{A}}\right]=\mathrm{tanh}\left(\frac{L_{\mathrm{M,}i}^{\mathrm{A}}}{2}\right)\cdot e^{j\varphi_{i}}\label{eq:SoftSymMap}
\end{equation}
based on, e.g., the feedback \textit{extrinsic} knowledge from the
single user channel decoder $L_{\mathrm{M,}i}^{\mathrm{A}}$. At the
first iteration, the feedbacks from channel decoders are $L_{\mathrm{M,}i}^{\mathrm{A}}=0$.
The feedback procedure, i.e., message passing from channel decoder
to the MUD, will be discussed later. For an arbitrary user $j$, the
output of the MUD after the SoIC can be written as
\begin{align}
y_{j} & =y-{\displaystyle \sum_{i=1,i\ne j}^{N}\sqrt{P_{i}}h_{i}\hat{x}_{i}}\label{eq:MUDSoIC}\\
 & =\sqrt{P_{j}}h_{j}\widetilde{x}_{j}+{\displaystyle \sum_{i=1,i\ne j}^{N}\sqrt{P_{i}}h_{i}\left(\tilde{x}_{i}-\hat{x}_{i}\right)}+n.\nonumber 
\end{align}

Then, each user starts its single user detection and decoding in parallel
based on the ``clean observations'' $y_{i}$. The (soft) demapper
computes the log-likelihood-ratio (LLR) of each bit while treating
the residual interference as noise. Assuming that the residual interference
and noise is Gaussian-distributed and let $\sigma_{\mathrm{I},j}^{2}$
denote the residual interference power for $j$th user, an approximation
of the true a posteriori LLR for BPSK can be computed according to
\begin{equation}
L_{\mathrm{M},j}^{\mathrm{E}}=4\sqrt{P_{j}}\frac{\mathrm{Re}\left\{ y_{j}\cdot h_{j}^{*}\cdot e^{-j\varphi_{j}}\right\} }{\sigma_{\mathrm{I},j}^{2}+\sigma_{n}^{2}}\label{eq:GALLR}
\end{equation}
where the noise variance $\sigma_{n}^{2}$, the random phase shifts
$\varphi_{j}$ and the channel coefficients $h_{j}$ are assumed to
be known to the receiver. The residual interference power after SoIC
can be estimated by
\begin{align}
\sigma_{\mathrm{I},j}^{2} & =\mathrm{E}\left[{\displaystyle \left|e^{-j\varphi_{j}}\sum_{i\ne j}\sqrt{P_{i}}h_{i}\left(\tilde{x}_{i}-\hat{x}_{i}\right)\right|^{2}}\right]\nonumber \\
 & =\sum_{i\ne j}P_{i}\left|h_{i}\right|^{2}\left(1-\mathrm{E}\left[\hat{x}_{i,m}^{2}\right]\right)\label{eq:InterferenceMeanEst}
\end{align}
where the expectation $\mathrm{E}\left[\cdot\right]$ is taken over
$m$. We assume that the interference term is Gaussian distributed,
which is approximately true provided that the number of users $N$
is large and the transmitted symbols are independent among users (central
limit theorem). 

Subsequently, the LLRs are deinterleaved (denoted by $\tilde{L}_{\mathrm{M},j}^{\mathrm{E}}=L_{\mathrm{R}\leftarrow M,j}^{\mathrm{A}}$
for the $j$th user which means the extrinsic message from MUD corresponds
to the \textit{a priori} knowledge of REP obtained by MUD) and sent
to a repetition decoder with the rate of $R_{r}=\frac{1}{d_{r}}$.
The extrinsic message of the $k$th information symbol of the repetition
code is the summation of the LLRs of all repeated symbols, given by
\begin{equation}
L_{\mathrm{D},j,k}^{\mathrm{A}}=L_{\mathrm{R\rightarrow D},j,k}^{\mathrm{E}}=\sum_{m=kd_{\mathrm{r}}}^{\left(k+1\right)d_{\mathrm{r}}-1}\tilde{L}_{\mathrm{M,}j,m}^{\mathrm{E}}
\end{equation}
where $m$ means the $m$th symbol of the repetition-coded codeword
and then it is forwarded to the LDPC decoder. Belief propagation (BP)
decoding can be performed by iterating between variable nodes (VNs)
and CNs of the LDPC code. The detailed decoding procedure will be
discussed in Sec. \ref{sec:EXIT-Analysis}. Let $L_{\mathrm{D},j,k}^{\mathrm{E}}=L_{\mathrm{R\leftarrow D,}i,m}^{\mathrm{A}}$
denote the extrinsic information about the $k$th symbol of an LDPC
codeword, i.e., output of the LDPC decoder. The extrinsic message
from the repetition decoder to the MUD can be written as 
\begin{equation}
L_{\mathrm{R\rightarrow M},j,k}^{\mathrm{E}}=\tilde{L}_{M,j,k}^{A}=L_{\mathrm{D},j,k}^{\mathrm{E}}+\sum_{m=nd_{\mathrm{r}},m\ne k}^{\left(n+1\right)d_{\mathrm{r}}-1}\tilde{L}_{\mathrm{M,}j,m}^{\mathrm{E}}.
\end{equation}
Afterwards, the LLRs are again interleaved and re-mapped to soft symbols
with \eqref{eq:SoftSymMap} for the MUD processing with \eqref{eq:MUDSoIC}.
In this way, multiple iterations can be carried out to reduce the
MAI. It is worth noting that the receiver needs the knowledge of active
users and their interleaver as well as phase scrambling for correct
decoding. 

Next, we discuss some key advantages of IDMA.
\begin{enumerate}
\item \textit{Low complexity}: Comparing the computation of LLR in \eqref{eq:GALLR}
and \eqref{eq:APPDec}, the SoIC-based MUD is quite simple and requires
only few multiplications and additions for variance estimation, interference
subtraction and LLR computation (see \cite{LipingIDMATWC06}). The
complexity becomes $O\left(M\cdot N\right)$ and many computations
can be carried out in parallel when compared to SIC. Particularly,
when all the users apply an LDPC code with the same parity check $\mathbf{H}$-matrix,
the coordination and code design among users can be dramatically reduced.
\item \textit{Parallelizable computation: }The $N$ single user channel
decoders can independently perform their decoding at the same time.
Since iterations between channel decoders and SoIC-MUD will be carried
out, the number of inner iterations within the channel decoders can
be kept very small. In fact, one iteration within the decoders suffices.
In contrast, SIC requires a considerable large number of inner iteration
within decoder for each user and often the operations are not parallelizable
due to its sequential processing. Hence, the latency may be reduced
in IDMA.
\item \textit{Asynchronous transmission}: The SoIC-MUD principle in \eqref{eq:MUDSoIC}
is applicable to asynchronous transmissions. For instance, \cite{KusumeIDMA12}
shows that the performance of IDMA is insensitive to different user
delays on the symbol-level. Without loss of generality, an asynchronous
transmission of two users' signals is illustrated in Fig.~\ref{fig:IDMAAsyn}.
The delay between users is denoted by $\tau_{d}$. Two or more consecutive
codewords shall be decoded within one common detection window and
interference cancellation can be performed within the window decoder
principle in \cite{CWMBwindowDecoder}.
\begin{figure}[tbh]
\begin{centering}
\includegraphics[width=0.95\columnwidth]{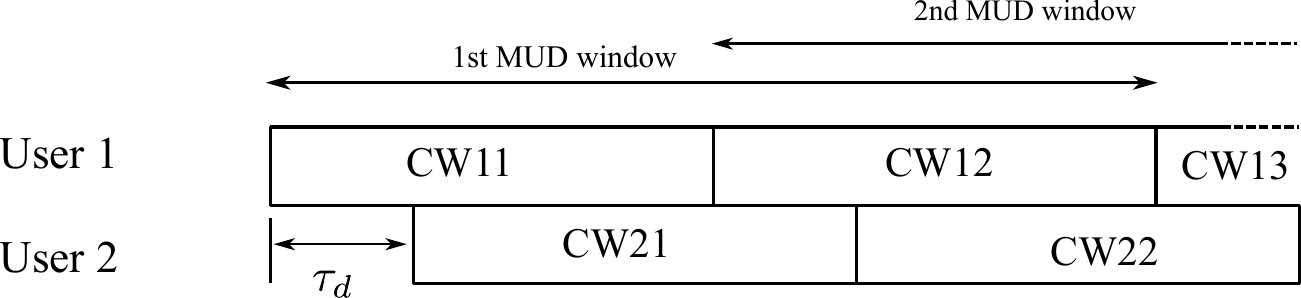}
\par\end{centering}
\caption{The iterative receiver processing of an asynchronous transmission
of two users' signals in IDMA; without loss of generality, the delay
$\tau_{d}$ between both users satisfies $0<\tau_{d}<\tau_{\mathrm{cw}}/2$
where $\tau_{\mathrm{cw}}$ is the time duration of one codeword.
\label{fig:IDMAAsyn}}
\end{figure}
\end{enumerate}

\section{\label{sec:EXIT-Analysis}Multiuser EXIT Analysis}

In order to optimize the LDPC degree profile, we elaborate on the
EXIT-chart approach in the context of IDMA by separately and/or jointly
considering the components of the multistage iterative receiver (MUD,
repetition decoder and LDPC decoder). For this, we illustrate the
graph representation of the BP based receiver in Fig. \ref{fig:GraphModelIDMARx}.
Furthermore, since we consider a system with a large number of users,
the messages being exchanged over the graph (LLRs) can be assumed
to be Gaussian distributed. We consider AWGN channels in the EXIT
analysis in the following, i.e., $h_{i,m}=1$, $\forall\,i,\,m$.

\begin{figure}[tb]
\begin{centering}
\begin{tikzpicture}[scale=0.7,xscale=-1, thick, MUDnode/.style={fill=myblue,draw,circle},   REPnode/.style={fill=mygreen,draw,circle},  VNDnode/.style={fill=myorgange,draw,circle}, CNDnode/.style={fill=myorgange,draw,rectangle}, every fit/.style={rectangle,draw,inner sep=2pt,text width=0.5cm},
>=latex]
\begin{scope}[start chain=going below,node distance=6mm] 
\foreach \i in {1,2,...,6}   
\node[MUDnode,on chain] (MUDn\i) [] {}; 
\end{scope}
\begin{scope}[xshift=4cm,yshift=-0.5cm,start chain=going below,node distance=4mm] 
\foreach \i in {1,2,...,3}   
\node[REPnode,on chain] (REPn\i) [] {}; 
\end{scope}
\begin{scope}[xshift=4cm,yshift=-4cm,start chain=going below,node distance=4mm] 
\foreach \i in {4,5,...,6}   
\node[REPnode,on chain] (REPn\i) [] {}; 
\end{scope}
\begin{scope}[xshift=5.8cm,yshift=-0.5cm,start chain=going below,node distance=4mm] 
\foreach \i in {1,2,...,3}   
\node[VNDnode,on chain] (VNDn\i) [] {}; 
\end{scope}
\begin{scope}[xshift=5.8cm,yshift=-4cm,start chain=going below,node distance=4mm] 
\foreach \i in {4,5,...,6}   
\node[VNDnode,on chain] (VNDn\i) [] {}; 
\end{scope}
\begin{scope}[xshift=7.8cm,yshift=-10mm,start chain=going below,node distance=4mm] 
\foreach \i in {1,2}   
\node[CNDnode,on chain] (CNDn\i) [] {}; 
\end{scope}
\begin{scope}[xshift=7.8cm,yshift=-45mm,start chain=going below,node distance=4mm] 
\foreach \i in {3,4}   
\node[CNDnode,on chain] (CNDn\i) [] {}; 
\end{scope}
\node [black,fit=(MUDn1) (MUDn6),label={[label distance=0.15cm]above: SoIC-MUD}] {};
\node [black,fit=(REPn1) (REPn3),label={[label distance=0.5cm]above: REP}] {};
\node (deinterleaver1) [black,fit=(REPn1) (REPn3),xshift=12.5mm] {$\pi_{1}$};
\node [black,fit=(REPn4) (REPn6)] {};
\node (deinterleaver2) [black,fit=(REPn4) (REPn6),xshift=12.5mm] {$\pi_{2}$};
\node [black,fit=(VNDn1) (VNDn3),label={[label distance=0.5cm]above: VND}] {};
\node [black,fit=(VNDn4) (VNDn6)] {};
\node [black,fit=(CNDn1) (CNDn2),label={[label distance=0.9cm]above: CND}] {};
\node [black,fit=(CNDn3) (CNDn4)] {};
\draw [myblue,thick,dashed] ($(deinterleaver1.north east) + (-0.2,0.1)$) rectangle ($(CNDn2.south west) + (0.5,-1)$);
\draw [myblue,thick,dashed] ($(deinterleaver2.north east) + (-0.2,0.1)$) rectangle ($(CNDn4.south west) + (0.5,-1)$);
\node [myblue, draw=none] at ($(CNDn1.north) + (-0.1,0.5)$) {User 1};
\node [myblue, draw=none] at ($(CNDn3.north) + (-0.1,0.5)$) {User 2};
\draw[<-] (MUDn1) -- node[xshift=0.1cm,above] {$y_1$}  +(-1.5,0); 
\draw[<-] (MUDn2) -- node[xshift=0.1cm,above] {$y_2$}  +(-1.5,0); 
\draw[<-] (MUDn3) -- +(-1.5,0); 
\draw[<-] (MUDn4) -- +(-1.5,0); 
\draw[<-] (MUDn5) -- node[xshift=0.15cm,above] {$y_m$}  +(-1.5,0); 
\draw[<-] (MUDn6) -- +(-1.5,0); 

\draw (MUDn1) -- ($(deinterleaver1.east)+(0,1)$); 
\draw (MUDn1) -- ($(deinterleaver2.east)+(0,0.8)$);
\draw (MUDn2) -- ($(deinterleaver1.east)+(0,0.6)$); 
\draw (MUDn2) -- ($(deinterleaver2.east)+(0,0.6)$);
\draw (MUDn3) -- ($(deinterleaver1.east)+(0,0.2)$); 
\draw (MUDn3) -- ($(deinterleaver2.east)+(0,0.2)$);
\draw (MUDn4) -- ($(deinterleaver1.east)+(0,-0.2)$); 
\draw (MUDn4) -- ($(deinterleaver2.east)+(0,-0.2)$);
\draw (MUDn5) -- ($(deinterleaver1.east)+(0,-0.4)$); 
\draw (MUDn5) -- ($(deinterleaver2.east)+(0,-0.4)$);
\draw (MUDn6) -- ($(deinterleaver1.east)+(0,-0.6)$); 
\draw (MUDn6) -- ($(deinterleaver2.east)+(0,-0.6)$);

\draw (REPn1) -- (deinterleaver1);
\draw (REPn1) -- ($(deinterleaver1.west)+(0,0.5)$);
\draw (REPn2) -- (deinterleaver1);
\draw (REPn2) -- ($(deinterleaver1.west)+(0,-0.5)$);
\draw (REPn3) -- (deinterleaver1);
\draw (REPn3) -- ($(deinterleaver1.west)+(0,-1)$);
\draw (REPn4) -- (deinterleaver2);
\draw (REPn4) -- ($(deinterleaver2.west)+(0,0.5)$);
\draw (REPn5) -- (deinterleaver2);
\draw (REPn5) -- ($(deinterleaver2.west)+(0,-0.5)$);
\draw (REPn6) -- (deinterleaver2);
\draw (REPn6) -- ($(deinterleaver2.west)+(0,-1)$);

\draw (REPn1) -- (VNDn1);
\draw (REPn2) -- (VNDn2);
\draw (REPn3) -- (VNDn3);
\draw (REPn4) -- (VNDn4);
\draw (REPn5) -- (VNDn5);
\draw (REPn6) -- (VNDn6);

\draw (CNDn1) -- (VNDn1);
\draw (CNDn1) -- (VNDn2);
\draw (CNDn1) -- (VNDn3);
\draw (CNDn2) -- (VNDn1);
\draw (CNDn2) -- (VNDn2);
\draw (CNDn2) -- (VNDn3);
\draw (CNDn3) -- (VNDn4);
\draw (CNDn3) -- (VNDn5);
\draw (CNDn3) -- (VNDn6);
\draw (CNDn4) -- (VNDn4);
\draw (CNDn4) -- (VNDn5);
\draw (CNDn4) -- (VNDn6);

\draw [decorate,decoration={brace,amplitude=10pt,mirror,raise=4pt},yshift=0pt] ($(REPn6)+(0.5,-1)$) -- +(-5.5,0) node [black,midway,yshift=-0.8cm] {\footnotesize MUD+REP};
\draw [decorate,decoration={brace,amplitude=10pt,raise=4pt},yshift=0pt] ($(REPn6)+(1.5,-1)$) -- +(3.2,0) node [black,midway,yshift=-0.8cm] {\footnotesize LDPC decoder};
\node[draw=none,rotate=-90, right = 1.2cm of MUDn2] {from channel};

\draw[rotate=-45,dashed] ($(MUDn1)+(0.65,-0.1)$) ellipse (5pt and 15pt) node[anchor=north west, xshift=-3mm,yshift=2mm,above]{$d_{u}$};
\draw[rotate=45,dashed] ($(REPn1)+(-0.8,0.2)$) ellipse (4pt and 6pt) node[anchor=north east, xshift=0.1mm,yshift=1.5mm,above]{$d_{r}$}; 

\draw[thick] ($(MUDn2)+(1.5,-0.1)$) ellipse (2pt and 50pt) node[anchor=north east, xshift=-3mm,yshift=10mm,above]{$\mathbf{y}_{1}$};
\draw[thick] ($(MUDn5)+(1.5,0.3)$) ellipse (2pt and 50pt) node[anchor=north east, xshift=1mm,yshift=-17mm,above]{$\mathbf{y}_{2}$};
\end{tikzpicture}
\par\end{centering}
\caption{Graph representation of a BP-based IDMA receiver (an example for two
users).\label{fig:GraphModelIDMARx}}
\end{figure}

\subsection{SoIC-MUD nodes}

As indicated in Fig. \ref{fig:GraphModelIDMARx}, each received symbol
$y_{m},\,m\in\left[0,N_{\mathrm{CW}}-1\right]$ is represented by
a MUD node, where $N_{\mathrm{CW}}$ denotes the length of a (after
repetition and LDPC encoding) codeword. A MUD node is connected to
one channel observation and $N$ repetition nodes, each REP node belonging
to another user. Consider an arbitrary MUD node, the message mean
being passed from repetition node of user $j$ is $\mu_{_{\mathrm{R\rightarrow M},j}}$.
After the SoIC, the remaining interference power can be derived with
\eqref{eq:InterferenceMeanEst} as
\begin{equation}
\sigma_{\mathrm{I},j}^{2}=\sum_{i=1,i\ne j}^{N}P_{i}\cdot\phi\left(\mu_{_{R\rightarrow M,i}}\right)\label{eq:InterferenceVar}
\end{equation}
by assuming $\mathrm{E}\left[\left|h_{i,m}\right|^{2}\right]=1,\,\forall i$
and $\phi\left(\mu\right)$ is given by
\begin{align}
\phi\left(\mu\right) & =\mathrm{E}\left[\left|\tilde{x}_{i}-\hat{x}_{i}\right|^{2}\right]\label{eq:PhiFuncDef}\\
 & =1-\int_{-\infty}^{\infty}\frac{e^{-\frac{y^{2}}{2}}}{\sqrt{2\pi}}\mathit{\mathrm{tanh}}\left(\frac{\mu}{2}-\sqrt{\frac{\mu}{2}}y\right)dy
\end{align}
which denotes the minimum mean square error (MMSE) \cite{GuoSVMMSEmIITI05}
of the SoIC with BPSK signaling. Thus, the updated message mean $\mu_{_{M\rightarrow R},j}$
can be expressed as 
\begin{equation}
\mu_{_{M\rightarrow R,j}}=\frac{4P_{j}}{\sigma_{n}^{2}+{\textstyle {\displaystyle \sum_{i=1,i\ne j}^{N}}}P_{i}\cdot\phi\left(\mu_{_{R\rightarrow M,i}}\right)}.\label{eq:MUDtoRepMean}
\end{equation}
It can be seen that the message mean depends on several factors, the
power distribution and the feedback message distribution. This complicates
code design as every user requires a matching code. In the following
(unless stated otherwise), we assume
\begin{align}
P_{1} & =P_{2}=\cdots=P_{N}=\frac{1}{N}\label{eq:SamePower}\\
\mu_{_{R\rightarrow M,1}} & =\mu_{_{R\rightarrow M,2}}=\cdots=\mu_{_{R\rightarrow M,N}}\eqqcolon\mu_{_{R\rightarrow M}}\label{eq:EQMsgRtoM}
\end{align}
and, thus
\begin{align}
\mu_{_{M\rightarrow R,1}} & =\mu_{_{M\rightarrow R,2}}=\cdots=\mu_{_{M\rightarrow R,N}}\eqqcolon\mu_{_{M\rightarrow R}}\label{eq:EQMsgMtoR}
\end{align}
so that \eqref{eq:MUDtoRepMean} can be simplified to 
\[
\mu_{_{M\rightarrow R}}=\frac{4}{N\sigma_{n}^{2}+\left(N-1\right)\cdot\phi\left(\mu_{_{R\rightarrow M}}\right)}.
\]
We note that the assumption in \eqref{eq:EQMsgRtoM} is only straightforward
for equal-power cases and every users applies the same LDPC and repetition
code. The extension to the unequal power case will be discussed in
Sec. \ref{sec:Beyond-Equal-Power}.

\subsection{Repetition nodes (REP)}

The repetition node has $d_{\mathrm{r}}$ edges to MUD nodes and one
edge to the VN of LDPC. Let $\mu_{_{V\rightarrow R}}$ denote the
message mean of LLRs of the VN decoder (VND), then we obtain the message
mean from the repetition node to MUD and VND as 
\begin{align}
\mu_{_{R\rightarrow M}} & =\left(d_{r}-1\right)\cdot\mu_{_{M\rightarrow R}}+\mu_{_{V\rightarrow R}}\label{eq:RepToMUD}\\
\mu_{_{R\rightarrow V}} & =d_{r}\cdot\mu_{_{M\rightarrow R}}\label{eq:RepToVND}
\end{align}
Upon convergence of the iterative processing between the SoIC-MUD
and the repetition nodes, we obtain 
\begin{equation}
\bar{\mu}_{_{M\rightarrow R}}=\frac{4}{N\sigma_{n}^{2}+\left(N-1\right)\phi\left(\left(d_{r}-1\right)\bar{\mu}_{_{M\rightarrow R}}+\mu_{_{V\rightarrow R}}\right)}.\label{eq:MUDRepConvMu}
\end{equation}
Note that solving the above equation \eqref{eq:MUDRepConvMu} for
$\bar{\mu}_{_{M\rightarrow R}}$ can capture the iterative behaviour
of the combined SoIC-MUD (considering a sufficient number of inner
iterations). 

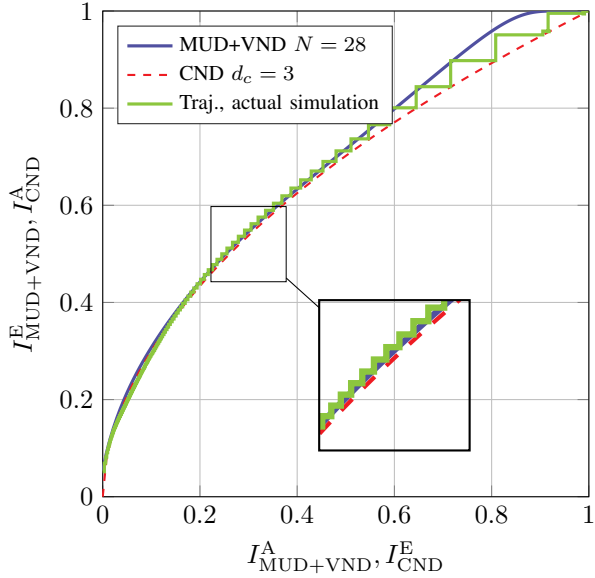
\begin{figure}[t]
\begin{centering}
\begin{tikzpicture} [spy using outlines=
	{magnification=2, connect spies}]
\begin{axis}[
width=\linewidth,
height=\linewidth,
xmajorgrids,
yminorticks=true,
ymajorgrids,
yminorgrids,
legend pos=north west,        
legend style={legend cell align=left,align=left,draw=white!15!black, font=\footnotesize},
xlabel={$I_{\mathrm{MUD+VND}}^{\mathrm{A}}, I_{\mathrm{CND}}^{\mathrm{E}}$},
ylabel={$I_{\mathrm{MUD+VND}}^{\mathrm{E}}, I_{\mathrm{CND}}^{\mathrm{A}}$},
mark size=1.5pt,
xmin=0,
xmax=1,
ymin=0,
ymax=1
]	

\addplot[color= myblue, very thick,each nth point={10}] table [x index=0,y index=1] {tikz/Data/EXIT_LDPC_OnlyCode_40dB_2_MI.dat};
\addlegendentry{MUD+VND $N=28$}

\addplot[color= rot, thick,dashed,each nth point={10}] table [x index=0,y index=2] {tikz/Data/EXIT_LDPC_OnlyCode_40dB_2_MI.dat};
\addlegendentry{CND $d_{c}=3$}

\addplot[color= apfelgruen, very thick] table [x index=0,y index=1] {tikz/Data/EXIT_LDPC_OnlyCode_40dB_1_MI.dat};
\addlegendentry{Traj., actual simulation}

\coordinate (spypoint1) at (axis cs:0.3,0.52);
\coordinate (magnifyglass1) at (axis cs:0.6,0.25);
\end{axis}

\spy [black,width=2cm,height=2cm] on (spypoint1) in node [fill=none] at (magnifyglass1);
\end{tikzpicture}
\par\end{centering}
\caption{EXIT-chart along with simulated trajectories of MUD and REP in an
``uncoded'' IDMA system with $N=32$ users and $\gamma_{s,\mathrm{mu}}=\unit[40]{dB}$;
all users have the same power $P_{i}=\frac{1}{N}$; the MAC system
is mainly interference-limited (see. Sec.~\ref{subsec:Operation-Regime})
and the repetition code is particularly effective in this operation
regime (compare to Fig. \ref{fig:Optimized-LDPC-Code}). \label{fig:EXIT-Chart-of-repetition-coded}}
\end{figure}

If no LDPC code is available, i.e., $\mu_{_{V\rightarrow R}}=0\,\forall\,i$,
the ``uncoded'' IDMA performance can be evaluated by the above message
passing mechanism. In Fig. \ref{fig:EXIT-Chart-of-repetition-coded},
the EXIT-chart is shown for a purely repetition-coded IDMA system
with $N=32$ users at a multi-user SNR of $\gamma_{s,\mathrm{mu}}=10\log_{10}\frac{\sum_{i=1}^{N}P_{i}}{\sigma_{n}^{2}}=\unit[40]{dB}$.

All the EXIT-curves are based on directly evaluating \eqref{eq:MUDtoRepMean}
and \eqref{eq:RepToMUD}, for instance, the SoIC-MUD EXIT-curve is
computed by 
\begin{equation}
I_{\mathrm{MUD}}^{E}=J\left(\frac{4}{N\sigma_{n}^{2}+\left(N-1\right)\cdot\phi\left(J^{-1}\left(I_{\mathrm{MUD}}^{A}\right)\right)}\right)\label{eq:MUDextEXIT}
\end{equation}
where the $J\left(\mu\right)$-function is given by 
\[
J\left(\mu\right)=1-\int_{-\infty}^{\infty}\frac{e^{-\frac{\left(y-\mu\right)^{2}}{4\mu}}}{\sqrt{4\pi\mu}}\log_{2}\left(1+e^{-y}\right)dy
\]
and its inverse $J^{-1}\left(I\right)$ can be approximated as in
\cite{FBrannstromEXITJapproxTIT}. The trajectories for the repetition
rate $R_{r}=\frac{1}{9}$ and $\frac{1}{12}$ obtained by numerical
simulations match very well with the EXIT-curve predictions. Small
deviations are still present in the high a priori knowledge region,
i.e., large $J\left(\mu_{_{R\rightarrow M}}\right)$, due to the GA.

\subsection{VN and CN}

The message exchange and update between VND and CN decoder (CND) is
well known \cite{tenBrinkLDPCEXITTC04}. For the sake of completeness,
we review and incorporate it into IDMA systems. Denote $\lambda_{i}$
and $\rho_{j}$ as the fraction of edges connected to VNs and CNs
with degrees $i$ and $j$, respectively. The message means $\mu_{_{V\rightarrow C}}^{i}$
and $\mu_{_{C\rightarrow V}}^{j}$, i.e., the mean of message from
CN (or VN) to VN (or CN) of degree $i$ (or $j$) respectively, are
given by 
\[
\mu_{_{V\rightarrow C}}^{i}=\mu_{_{R\rightarrow V}}+\left(i-1\right)\cdot{\displaystyle \sum_{j=2}^{c_{\mathrm{max}}}}\rho_{j}\mu_{_{C\rightarrow V}}^{j}
\]
where $c_{\mathrm{max}}$ denotes the maximum degree of CNs, and 
\[
\mu_{_{C\rightarrow V}}^{j}=\phi^{-1}\left(1-\left(1-\sum_{i=2}^{v_{\mathit{\mathrm{max}}}}\lambda_{i}\phi\left(\mu_{_{V\rightarrow C}}^{i}\right)\right)^{j-1}\right)
\]
where $v_{\mathrm{max}}$ denotes the maximum degree of VNs and the
channel observation of the VND is now provided by the REP. Note that
the $\phi\left(\mu\right)$ function given in \eqref{eq:PhiFuncDef}
in this paper is equivalent to that in \cite[Def. 1]{910580} by means
of change of variables. Furthermore, we consider \textit{check-regular}
LDPC codes, i.e., $\rho_{j}=1$ for a given $d_{c}=j$. The VND also
passes a message to the connected repetition node, given by 
\[
\mu_{_{V\rightarrow R}}^{i}=i\cdot{\displaystyle \sum_{j=2}^{c_{\mathrm{max}}}}\rho_{j}\mu_{_{C\rightarrow V}}^{j}.
\]

\section{\label{sec:Degree-Profile-Optimization}Universal Code Design}

The detection and decoding of signals in the MAC is challenging not
only due to the noise but also due to the MAI. The code design for
the MAC therefore depends also on the relation between the MAI and
the noise. We briefly discuss the code design for different operation
regimes of a MAC in the following and then present a flexible code
design for a dynamic MAC system with varying number of users.

\subsection{Operation Regime\label{subsec:Operation-Regime}}

In a multi-user transmission, the achievable sum-rate is determined
by the multi-user SNR given by 
\[
\gamma_{\mathrm{s,mu}}=\frac{\sum_{i=1}^{N}{\displaystyle P_{i}\cdot\mathrm{E}\left[\left|h_{i}\right|^{2}\right]}}{\sigma_{n}^{2}}
\]
and the ultimate achievable rate of each user is determined by the
single-user SNR, e.g., for the $i$th user we have 
\[
\gamma_{\mathrm{s,su}}\left(i\right)=\frac{{\displaystyle P_{i}\cdot\mathrm{E}\left[\left|h_{i}\right|^{2}\right]}}{\sigma_{n}^{2}}.
\]
However, this single-user SNR is only attainable when all other users'
signals are perfectly decoded and canceled out. In the iterative detection
and decoding procedure, the performance is limited by both noise and
MAI, i.e., 
\[
\frac{{\displaystyle P_{i}\cdot\mathrm{E}\left[\left|h_{i}\right|^{2}\right]}}{\underset{\sigma_{I,i}^{2}}{\underbrace{{\displaystyle \sum_{j=1,j\ne i}^{N}P_{j}\cdot\mathrm{E}\left[\left|h_{j}\right|^{2}\right]}}}+\sigma_{n}^{2}}\leqslant\gamma_{\mathrm{s,su}}\left(i\right)\leqslant\frac{{\displaystyle P_{i}\cdot\mathrm{E}\left[\left|h_{i}\right|^{2}\right]}}{\sigma_{n}^{2}}.
\]

\begin{figure}[tb]
\begin{centering}
\tikzfading[name=fade down1,top color=green!0, bottom color=red!100]

\tikzfading[name=fade up1,top color=red!100, bottom color=green!0]

\tikzfading[name=fade down2,top color=green!0, bottom color=blue!100]

\tikzfading[name=fade up2,top color=blue!100, bottom color=green!0]

\begin{tikzpicture} 
\begin{axis}[
width=\linewidth,
height=\linewidth,
xmajorgrids,
ymajorgrids,
legend pos=north west,        
xlabel={SNR $\gamma_{\mathrm{s,mu}}=\frac{1}{\sigma_{n}^{2}}$ in dB},
ylabel={INR $\gamma_{\mathrm{INR}}=\frac{d_{u}-1}{d_{u}\cdot\sigma_{n}^{2}}$ in dB},
mark size=10pt,
xmin=-10,
xmax=10,
ymin=-10,
ymax=10,
]	
\node[fill=red, circle,inner sep=0.7pt,minimum size=1pt] (p0)  at (axis cs: -10, -10.1379) {};
\node[draw=none] (p01)  at (axis cs: -6, -6.1379) {};
\node[draw=none] (p1) at (axis cs: 0.35,0.2121) {};
\node[draw=none] (p11) at (axis cs: -0.35,-0.4879) {};
\node[draw=none] (p111) at (axis cs: -2.35,-2.4879) {};
\node[draw=none] (p21)  at (axis cs: 6, 5.8621) {};
\node[fill=blue,circle,inner sep=0.7pt,minimum size=1pt] (p2)  at (axis cs: 10.1, 9.9621) {};

\path[path fading=fade up1, ultra thick, draw=red] (p0) -- (p1);
\path[path fading=fade down1, ultra thick, draw=green] (p1) -- (p0);
\path[path fading=fade up1, ultra thick, draw=red] (p0) -- (p01);
\node[color=red] at (axis cs: -4,-8) {noise-limited};

\path[path fading=fade down2, ultra thick, draw=blue] (p11) -- (p2);
\path[path fading=fade up2, ultra thick, draw=green] (p2) -- (p11);
\path[path fading=fade down2, ultra thick, draw=blue] (p21) -- (p2);
\node[color=green] at (axis cs: -2, 2) {MAI$\approx$noise};

\node[color=blue] at (axis cs: 4.5, 8) {MAI-limited};

\end{axis}

\end{tikzpicture}
\par\end{centering}
\caption{Illustration of the three operation regimes, i.e., noise-limited,
MAI-limited, MAI and noise-limited; equal user power is assumed, i.e.,
$P_{1}=P_{2}=\cdots=P_{N}=\frac{1}{N}$. For example, $\sigma_{I,i}^{2}\approx\sigma_{n}^{2}$
may be assumed when the ratio $\frac{\sigma_{I,i}^{2}}{\sigma_{n}^{2}}$
is within $\pm3\,\mathrm{dB}$.\label{fig:IIustration-of-the}}
\end{figure}
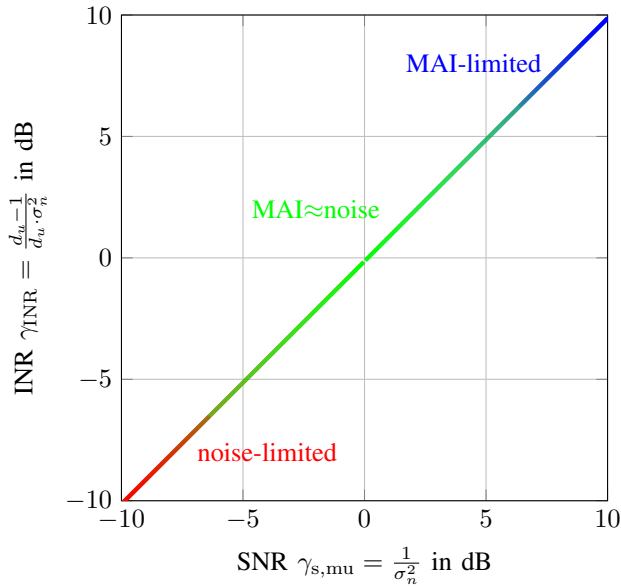

Depending on the noise variance and MAI, we can distinguish three
regimes of operation:
\begin{enumerate}
\item The noise-limited case, with the noise power dominating, i.e., $\sigma_{n}^{2}\gg\sigma_{I,i}^{2}$.
Single user code design is sufficient since the effect of removing
the MAI is marginal.
\item The MAI-limited case, with MAI dominating, i.e., $\sigma_{I,i}^{2}\gg\sigma_{n}^{2}$.
Orthogonal approaches and/or repetition coding are sufficient. We
will see an example of this regime of operation in the following section.
\item The MAI and noise-limited case, with $\sigma_{I,i}^{2}\approx\sigma_{n}^{2}$.
This case is quite challenging and of most practical interest. This
paper mainly focuses on this case.
\end{enumerate}
Fig. \ref{fig:IIustration-of-the} illustrates the three regimes of
operation assuming equal transmit power per user. The interference-to-noise
ratio (INR) case can then be written as
\[
\gamma_{_{\mathrm{INR}}}=\frac{N-1}{N}\cdot\gamma_{s,\mathrm{mu}}.
\]

\subsection{LDPC Code Only}

The LDPC degree profile, i.e., the coefficients $\lambda_{i}$ and
$\rho_{j}$, can be optimized to match (curve fitting) the EXIT-curve
of the inner SoIC-MUD. The optimization approach is summarized in
Algorithm \ref{alg:LDPC-code-design-1}. The repetition code is excluded
at first in this subsection, i.e., $d_{r}=1$ and all redundancy is
devoted to the LDPC code. In this case, the MUD, repetition and VND
nodes can be merged into one effective MUD-VND node, and message passing
is performed solely between CND and MUD-VND nodes.
\begin{algorithm}[tbh]
\textbf{Input}: Number of users $N$, maximum VN degree $v_{\mathrm{max}}$,
maximum CN degree $c_{\mathrm{max}}$, maximum repetition factor $r_{\mathrm{max}}$,
noise variance $\sigma_{n}^{2}$

\textbf{Output}: Optimal VN degree distribution $\boldsymbol{\mathbf{\lambda}}_{\mathrm{opt}}$,
optimal VN degree distribution $\boldsymbol{\mathbf{\rho}}_{\mathrm{opt}}$,
optimal repetition factor $d_{r,\mathrm{opt}}$, maximal code-rate
$R_{\mathrm{max}}$

for $d_{r}=1$ to $r_{\mathrm{max}}$

\ for $d_{c}=2$ to $c_{\mathrm{max}}$

\qquad{}Solve LP: 

\begin{align}
\underset{\mathbf{\lambda}_{\mathrm{opt}}}{\mathrm{max}}\: & {\displaystyle \sum_{i=2}^{v_{\mathrm{max}}}\frac{\lambda_{i}}{i}}\nonumber \\
\mathrm{s.t.} & \lambda_{i}\geq0\nonumber \\
 & {\displaystyle \sum_{i=2}^{v_{\mathrm{max}}}}\lambda_{i}=1\nonumber \\
 & {\displaystyle 1-\sum_{i=2}^{v_{\mathit{\mathrm{max}}}}\lambda_{i}\phi\left(d_{r}\cdot\bar{\mu}_{M\rightarrow R}\left(i\cdot\mu\right)+\left(i-1\right)\cdot\mu\right)}\nonumber \\
 & >\sqrt[d_{c}-1]{1-\phi\left(\mu\right)}\label{eq:StabCon2}\\
 & \lambda_{2}\leq\frac{e^{\frac{1}{N\sigma_{n}^{2}}}}{d_{c}-1}\label{eq:StabCond-1}
\end{align}

\qquad{}Compute $R=\frac{1}{d_{r}}\cdot\left(1-\frac{1}{d_{c}\sum_{i=2}^{v_{\mathrm{max}}}{\displaystyle \frac{\lambda_{i}}{i}}}\right)$

\ end

end

Select the maximum $R_{\mathrm{max}}$ and output the corresponding
$d_{r,\mathrm{opt}}$, $\boldsymbol{\mathbf{\lambda}}_{\mathrm{opt}}$
and $\mathbf{\boldsymbol{\rho}}_{\mathrm{opt}}$.

\caption{LDPC code design based on EXIT-charts with and without repetition
code.\label{alg:LDPC-code-design-1}}
\end{algorithm}
 As shown in Appendix~\ref{subsec:Proof-of-1}, the stability condition
in \eqref{eq:StabCond-1} is derived. The LDPC code shall be optimized
and used to mitigate the MAI and noise disturbance. Note that the
initial SINR (no iteration) can be very small. For instance, the SINRs
are $\unit[-18]{dB}$ and $\unit[-14.7]{dB}$ for $\gamma_{s,\mathrm{mu}}=\unit[0]{dB}$
and $\unit[40]{dB}$ (both cases with $N=32$ users), respectively.

\begin{figure}[tbh]
\begin{centering}
\begin{tikzpicture} [spy using outlines=
	{magnification=2, connect spies}]
\begin{axis}[
width=\linewidth,
height=\linewidth,
xmajorgrids,
yminorticks=true,
ymajorgrids,
yminorgrids,
legend pos=north west,        
legend style={legend cell align=left,align=left,draw=white!15!black, font=\footnotesize},
xlabel={$I_{\mathrm{MUD+VND}}^{\mathrm{A}}, I_{\mathrm{CND}}^{\mathrm{E}}$},
ylabel={$I_{\mathrm{MUD+VND}}^{\mathrm{E}}, I_{\mathrm{CND}}^{\mathrm{A}}$},
mark size=1.5pt,
xmin=0,
xmax=1,
ymin=0,
ymax=1
]	

\addplot[color= myblue, very thick,each nth point={10}] table [x index=0,y index=1] {tikz/Data/EXIT_LDPC_OnlyCode_40dB_2_MI.dat};
\addlegendentry{MUD+VND $N=28$}

\addplot[color= rot, thick,dashed,each nth point={10}] table [x index=0,y index=2] {tikz/Data/EXIT_LDPC_OnlyCode_40dB_2_MI.dat};
\addlegendentry{CND $d_{c}=3$}

\addplot[color= apfelgruen, very thick] table [x index=0,y index=1] {tikz/Data/EXIT_LDPC_OnlyCode_40dB_1_MI.dat};
\addlegendentry{Traj., actual simulation}

\coordinate (spypoint1) at (axis cs:0.3,0.52);
\coordinate (magnifyglass1) at (axis cs:0.6,0.25);
\end{axis}

\spy [black,width=2cm,height=2cm] on (spypoint1) in node [fill=none] at (magnifyglass1);
\end{tikzpicture}
\par\end{centering}
\caption{EXIT-chart of an optimized LDPC code with Algorithm \ref{alg:LDPC-code-design-1}
at SNR of $\gamma_{s,\mathrm{mu}}=\unit[40]{dB}$ with $N=28$ and
$R_{c}=0.1068$; the system operates in the MAI-limited regime; the
achievable rate with optimized LDPC code is smaller than that with
simple a repetition code (compare with Fig. \ref{fig:EXIT-Chart-of-repetition-coded}).\label{fig:Optimized-LDPC-Code}}
\end{figure}
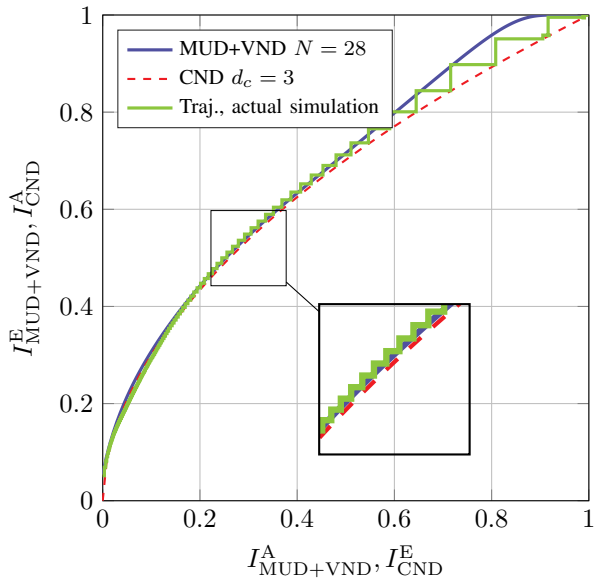

In Fig. \ref{fig:Optimized-LDPC-Code}, we show the EXIT-chart of
the optimized LDPC code of rate $R_{c}=0.1068$ at $\gamma_{s,\mathrm{mu}}=\unit[40]{dB}$
and $N=32$. We limit the maximum VN and CN degrees to $v_{\mathrm{max}}=320$
and $c_{\mathrm{max}}=64$ to enable feasible finite length code constructions\footnote{It turns out that not all degrees, particularly the high degrees,
are needed.}. In the simulation, random but user-specific interleavers are used
and the LDPC codeword length is set to $L_{\mathrm{CW}}=10^{4}$.
Throughout the paper, the parity check matrices $\mathbf{H}$ are
randomly generated according to the optimized LDPC degree profile,
and then the matrix with the largest average girth is selected for
numerical simulation.

It turns out that convergence is achieved with $N=28$ users when
fixing the multi-user SNR $\gamma_{s,\mathrm{mu}}$ to $\unit[40]{dB}$.
We treat the SoIC-MUD nodes and the VND of the LDPC as one component,
and the CND as another component. The analytical EXIT-functions are
depicted by the solid blue and dashed red curves, respectively. For
each iteration, the mutual information of the exchanged messages between
the two components is simulated. The average decoding trajectory (denoted
by green staircase, obtained from simulation) is shown in Fig. \ref{fig:Optimized-LDPC-Code}
for $171$ iterations. It can be seen that the trajectory matches
quite well with the analytical EXIT-functions,  which verifies the
viability of the algorithm. 

The optimized LDPC code at such high SNR is not as efficient as the
simple repetition code both in terms of decoding complexity and achievable
sum-rate. In the repetition-coded system, a $R_{r}=$1/9 repetition
code is able to support 32 users and practically achieves error-free
decoding in 20 iterations (comparing Fig. \ref{fig:EXIT-Chart-of-repetition-coded}),
leading to the sum-rate of $R_{\mathrm{sum}}^{\mathrm{REP}}=NR_{r}=\unit[3.56]{bpcu}$.
With the optimized LDPC code, the achievable rate of $R_{\mathrm{sum}}^{\mathrm{LDPC}}=NR_{c}=\unit[2.99]{bpcu}$
is achieved with $171$ iterations. At such high SNR $\gamma_{s,\mathrm{mu}}=\unit[40]{dB}$,
the performance is mainly interference-limited. Interestingly, the
required minimal repetition code rate is $\frac{1}{9}$ while a spreading
factor of $16$ would be necessary in an orthogonal CDMA scheme. This
indicates that repetition codes are very efficient for interference
cancellation. In a more realistic SNR region, the noise sets the upper
limit on the MUD, i.e., $I_{\mathrm{E}}=J\left(\frac{1}{N\sigma_{n}^{2}}\right)$.
Therefore, an LDPC code is further required to mitigate the noise
disturbance, as described next. 

In Fig. \ref{fig:BER-over-EbN0}, the BER of the optimized LDPC code
at the target sum-rate of $R_{\mathrm{sum}}=N\frac{R_{c}}{d_{r}}=\unit[1]{bpcu}$
is shown. The interleaver depth is set to $\unit[d_{r}L_{\mathrm{CW}}]{bits}$
in the simulation to mitigate short cycles due to the superposition
of multiple users. It turns out that the performance depends, to a
great extent, on the number of users $N$. In the simulation, we set
the number of users to $N=30$ although the code is optimized for
$N=32$; this ``slack'' is to account for a possibly small girth
of the parity check matrix and any other finite length effects that
would slow down convergence. The corresponding GMAC capacity is
achieved at $-\unit[0.1]{dB}$, and the optimized code is more than
$\unit[6]{dB}$ away from the GMAC capacity. The reasons are manifold;
firstly, it is well known that LDPC code design does not yield satisfactory
performance at low code rates \cite{AndriyanovaLowRateCodeTC12};
secondly, the BP-graph is rather dense so that convergence is compromised.
We remark that the performance can be improved by increasing the interleaver
depth and/or constructing $\mathbf{H}$ matrices using more sophisticated
design methods such as progressive edge growth (PEG) \cite{PEG} and
methods in \cite{LivaStructuredLDPC}.

\subsection{Repetition code and LDPC}

The repetition code seems to be an effective means for supporting
interference cancellation. Thus, we extend the LDPC degree profile
optimization to a serially concatenated repetition code with $d_{r}>1$.
The procedure for the optimization of the degree profile involving
a repetition code is summarized in Algorithm \ref{alg:LDPC-code-design-1}.
If a target code rate $R_{c}$ is required, the Algorithm \ref{alg:LDPC-code-design-1}
can be recursively applied for different values of the noise variance
$\sigma_{n}^{2}$ (e.g., bisection method depending on the code-rate
output of the algorithm). Hereby, the MUD, repetition and VND nodes
are merged into one effective MUD-REP-VND node. Within this node,
a sufficient number of iterations is carried out so that the messages
between MUD and repetition node have converged. The converged message
mean, denoted by $\bar{\mu}_{M\rightarrow R}\left(i\cdot\mu\right)$,
depends on the incoming message from the connected VND $\mu_{V\rightarrow R}^{i}=i\cdot\mu$
and can be obtained by solving \eqref{eq:MUDRepConvMu}. 

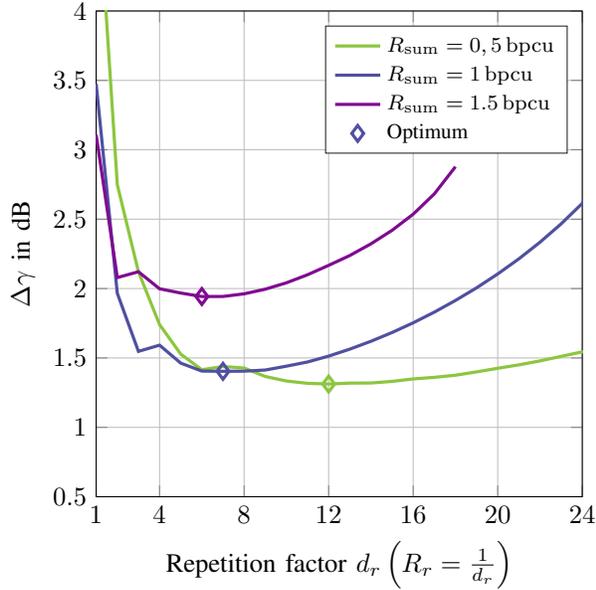
\begin{figure}[tbh]
\begin{centering}
\begin{tikzpicture} [spy using outlines=
	{magnification=2, connect spies}]
\begin{axis}[
width=\linewidth,
height=\linewidth,
xmajorgrids,
yminorticks=true,
ymajorgrids,
yminorgrids,
legend pos=north east,        
legend style={legend cell align=left,align=left,draw=white!15!black, font=\footnotesize},
xlabel={Repetition factor $d_{r} \left(R_{r}=\frac{1}{d_{r}}\right)$},
ylabel={$\Delta\gamma$ in dB},
xtick={1,4,8,12,16,20,24},
mark size=3pt,
xmin=1,
xmax=24,
ymin=0.5,
ymax=4
]	

\addplot[color= apfelgruen, very thick,each nth point={1}] table [x index=0,y index=1] {tikz/Data/Gap_Rate_0.5.dat};
\addlegendentry{$R_{\mathrm{sum}}=\unit[0,5]{bpcu}$}

\addplot[color= myblue, very thick,each nth point={1}] table [x index=0,y index=1] {tikz/Data/EbN0_Gap_rate1_Rep.dat};
\addlegendentry{$R_{\mathrm{sum}}=\unit[1]{bpcu}$}

\addplot[color= lila, very thick,each nth point={1}] table [x index=0,y index=1] {tikz/Data/Gap_Rate_1.5.dat};
\addlegendentry{$R_{\mathrm{sum}}=\unit[1.5]{bpcu}$}


\addplot[color= myblue, very thick, mark=diamond, only marks ] coordinates {(7, 1.4038)};
\addlegendentry{Optimum}
\addplot[color= lila, very thick, mark=diamond ] coordinates {(6.0000000e+00,  1.9425914e+00)};
\addplot[color= apfelgruen, very thick, mark=diamond ] coordinates {(1.2000000e+01,   1.3120996e+00)};


\end{axis}

\end{tikzpicture}
\par\end{centering}
\caption{Impact of the repetition code on the BP-threshold of the LDPC code
design and the (BP) gap-to-capacity $\Delta\gamma$ with $N=32$ users.\label{fig:Impact-of-repetition}}
\end{figure}

In Fig. \ref{fig:Impact-of-repetition}, we present the $\frac{\mathrm{E}_{b}}{N_{0}}$-gap
to the Shannon limit of the joint repetition and LDPC code optimization
for various repetition factors $d_{r}$ and a few target sum-rates,
where the gap is calculated as 
\[
\Delta\gamma=\frac{\zeta_{t}}{2^{R_{\mathrm{sum}}}-1}
\]
where $\zeta_{t}$ denotes the decoding threshold (required SNR) and
$R_{\mathrm{sum}}$ denotes the achieved sum rate of all users. 
 As can be observed, there exists an optimum repetition code rate
$\frac{1}{d_{r}}$ for the outer LDPC code rate $R_{c}$ which maximizes
the spectral efficiency. The optimum repetition factor $d_{r}$ is
marked in Fig. \ref{fig:Impact-of-repetition}. This also indicates
that repetition codes are efficient for interference cancellation,
since it is well known that repetition codes have no coding gain in
the single user case.
We note that the SoIC-MUD based on GA is sub-optimal for the finite
modulation. The loss incurred by the sub-optimality becomes significant
when the SNR increases. Thus, the $\frac{\mathrm{E}_{b}}{N_{0}}$-gap
is larger when the target sum-rate increases. 

We select the target sum rate of $R_{\mathrm{sum}}=\unit[1]{bpcu}$
whose GMAC capacity is achieved at $\frac{E_{b}}{N_{0}}=\gamma_{s,\mathrm{mu}}=\unit[0]{dB}$.
In this case, the MAI and the noise have approximately the same power
such that the interference cancellation and error correction are of
comparable importance.
\begin{table}[tbh]
\begin{centering}
\caption{LDPC parameters \label{tab:LDPC-parameters}}
\par\end{centering}
\centering{}%
\begin{tabular}{|c|c|}
\hline 
\multicolumn{2}{|c|}{$N=32$ and $R_{\mathrm{sum}}=N\frac{R_{c}}{d_{r}}=\unit[1]{bpcu}$}\tabularnewline
\hline 
\hline 
\textbf{$d_{r}=2$} & $d_{r}=4$\tabularnewline
\hline 
$R_{c}=0.0625$ & $R_{c}=0.125$\tabularnewline
\hline 
$R_{\mathrm{tot}}=0.03125$ & $R_{\mathrm{tot}}=0.03125$\tabularnewline
\hline 
$d_{c}=3,\rho_{3}=1$ & $d_{c}=3,\rho_{3}=1$\tabularnewline
\hline 
$\lambda_{2}=0.5252$ & $\lambda_{2}=0.5231$\tabularnewline
$\lambda_{3}=0.2112$ & $\lambda_{3}=0.3187$\tabularnewline
$\lambda_{9}=0.1030$ & $\lambda_{12}=0.1582$\tabularnewline
$\lambda_{10}=0.0915$ & \tabularnewline
$\lambda_{35}=0.0587$ & \tabularnewline
$\lambda_{36}=0.0104$ & \tabularnewline
 & \tabularnewline
\hline 
\hline 
$d_{r}=6$ & $d_{r}=8$\tabularnewline
\hline 
$R_{c}=0.1875$ & $R_{c}=0.25$\tabularnewline
\hline 
$R_{\mathrm{tot}}=0.03125$ & $R_{\mathrm{tot}}=0.03125$\tabularnewline
\hline 
$d_{c}=4,\rho_{3}=1$ & $d_{c}=5,\rho_{3}=1$\tabularnewline
\hline 
$\lambda_{2}=0.3480$ & $\lambda_{2}=0.2610$\tabularnewline
$\lambda_{3}=0.3450$ & $\lambda_{3}=0.3505$\tabularnewline
$\lambda_{15}=0.1829$ & $\lambda_{16}=0.2526$\tabularnewline
$\lambda_{16}=0.0939$ & $\lambda_{17}=0.0361$\tabularnewline
$\lambda_{49}=0.0302$ & $\lambda_{70}=0.0778$\tabularnewline
 & $\lambda_{71}=0.0220$\tabularnewline
\hline 
\end{tabular}
\end{table}
 The INR can be calculated to $\gamma_{_{\mathrm{INR}}}=\unit[-0.1379]{dB}$
with $N=32$. The optimized LDPC code parameters\footnote{The parity check $\mathbf{H}$-matrices used for the simulation in
this paper can be found in \cite{GithupPCM}} are given in Tab. \ref{tab:LDPC-parameters}.  

\begin{figure}[tbh]
\begin{centering}
\begin{tikzpicture} [spy using outlines=
	{magnification=2, connect spies}]
\begin{axis}[
width=\linewidth,
height=\linewidth,
xmajorgrids,
yminorticks=true,
ymajorgrids,
yminorgrids,
legend pos=north west,        
legend style={ at={(0.22,1)},legend cell align=left,align=left,draw=white!15!black, font=\footnotesize},
xlabel={$\frac{E_{b}}{N_{0}}$ in dB},
ylabel={BER},
ymode=log,
mark size=1.5pt,
xmin=-0.2,
xmax=6,
ymin=1e-4,
ymax=1
]	

\addplot[color= black, thick,dashed ] coordinates {(-0.1, 1e-4) (-0.1,1e-2)};
\addlegendentry{GMAC capacity}

\addplot[color= orange, mark=o, very thick,each nth point={1}] table [x index=0,y index=1] {tikz/Data/BER_30user_Rep_1_snrs.dat};
\addlegendentry{$d_{r}=1, R_{c}=0.03125$, AWGN}

\addplot[color= violettblau, mark=x, very thick, each nth point={1}] table [x index=0,y index=1] {tikz/Data/BER_30user_Rep_2_snrs.dat};
\addlegendentry{$d_{r}=2, R_{c}=0.0625$, AWGN}

\addplot[color= lila, mark=x, very thick,each nth point={1}] table [x index=0,y index=1] {tikz/Data/BER_30user_Rep_4_snrs.dat};
\addlegendentry{$d_{r}=4, R_{c}=0.125$, AWGN}

\addplot[color= dunkelgrau, mark=+, very thick,each nth point={1}] table [x index=0,y index=1] {tikz/Data/BER_30user_Rep_4_Ray_snrs.dat};
\addlegendentry{$d_{r}=4, R_{c}=0.125$, Rayleigh}

\addplot[color= apfelgruen, mark=square, very thick,each nth point={1}] table [x index=0,y index=1] {tikz/Data/BER_30user_SU_Gauss_Avg_Rep_4_snrs.dat};
\addlegendentry{$d_{r}=4, R_{c}=0.125$, SU code}







\end{axis}

\end{tikzpicture}
\par\end{centering}
\caption{BER comparison for $R_{\mathrm{sum}}=\unit[0.9375]{bpcu}$ with $N=$30
users and the repetition factors $d_{r}=1,2$ and $4$; the overall
per-user code-rate is $R_{\mathrm{tot}}=\frac{R_{c}}{d_{r}}=0.03125$;
the codes are optimized for $N=$32 users except for the single-user
reference (``SU code''), denoted by the green curve.\label{fig:BER-over-EbN0}}
\end{figure}
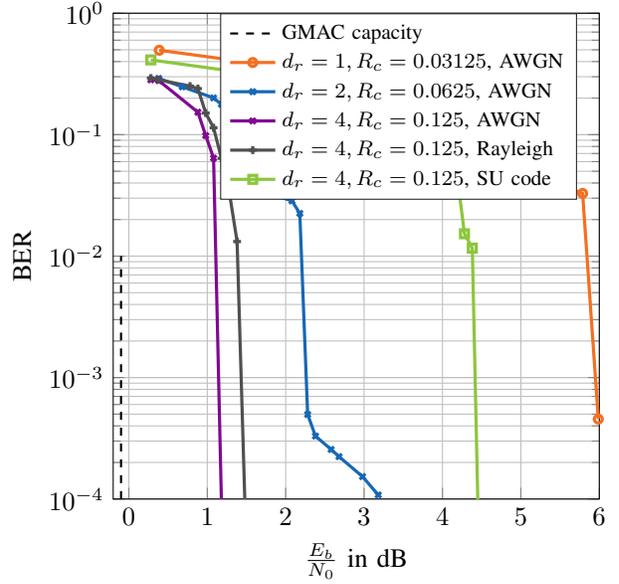

Fig. \ref{fig:BER-over-EbN0} shows the BER for different repetition
factors $d_{r}$. The optimized code with $d_{r}=4$ achieves the
BER of $10^{-4}$ at the $\frac{E_{b}}{N_{0}}$ of $\unit[1.18]{dB}$,
which is only $\unit[1.28]{dB}$ away from the GMAC capacity (the
corresponding $\frac{E_{b}}{N_{0}}$ GMAC limit is $-\unit[0.1]{dB}$).
The performance gain through inclusion of a repetition code is obvious.
Further increase of the repetition factor does not offer significant
gains (not shown). Moreover, the Rayleigh fading channel is considered
for $d_{r}=4$. Only $\unit[0.2]{dB}$ loss is incurred compared to
the AWGN channel. This is due to the multi-user diversity in the multi-access
fading channels \cite{DTseBookFund} and the low-rate per user. To
validate the matching gain of the multi-user code design, i.e., the
performance gain of a dedicated LDPC code optimization for a multi-user
setup compared to an LDPC code optimized for a single-user setup,
we further simulate the performance of a single-user code of rate
$R_{c}=0.125$ which is optimized for the single-user (SU) AWGN channel.
To support $N=30$ users in IDMA, the optimized SU code is serially
concatenated with the REP code ($d_{r}=4$), yielding a matching gain
of more than $\unit[3]{dB}$.

\subsection{Repetition as User-load Equalizer}

The degree profile optimization and the finite-length code construction
of the LDPC code is computationally quite expensive particularly if
the repetition code is serially concatenated and/or more sophisticated
optimizations are considered (e.g., ``full'' density evolution and/or
irregular CNs). When the number of users varies in a system, usually,
the code would have to be re-optimized to achieve closer to MAC capacity,
or several pre-optimized LDPC codes have to be stored for each possible
scenario (e.g., amount of users, transmit power distribution, channel
condition). This limits the flexibility and practical usability of
multiple access schemes that are based on variants of superposition
coding, like IDMA. In the following we show how this issue can be
mitigated, working close to capacity for any user load using the \textit{same}
LDPC parity check matrix $\mathbf{H}$ by just adapting the repetition
code rate $R_{r}=\frac{1}{d_{r}}$.

Considering the LDPC code (the same for all users), we observe that
the ``channel observation'' of the VN is provided by the MUD+REP
detector and its message mean is given by $\mu_{_{R\rightarrow V}}$.
Allowing a sufficient number of iterations between the repetition
node and the MUD, we obtain the ``channel input'' to the VN of the
LDPC code by combining \eqref{eq:RepToVND} and \eqref{eq:MUDRepConvMu}
as
\begin{align*}
\mu_{_{R\rightarrow V}} & =\frac{d_{\mathrm{r}}}{N}\frac{4}{\sigma_{n}^{2}+\left(1-\frac{1}{N}\right)\phi\left(\left(d_{r}-1\right)\bar{\mu}_{_{M\rightarrow R}}+\mu_{_{V\rightarrow R}}\right)}.
\end{align*}
This ``channel input'' to the LDPC decoder depends on \textit{three}
system parameters, i.e., $N$, $d_{r}$ and $\sigma_{n}^{2}$. Note
that it also depends on the feedback information from the channel
decoder $\mu_{_{V\rightarrow R}}$, however, we do not consider it
as system parameter. In other words, the ``channel input'' is characterized
by a \textit{three dimensional} function, i.e., $\mu_{_{R\rightarrow V}}\left(N,d_{r},\sigma_{n}^{2}\right)$.
We prove that the dimensionality can be further reduced to $\mu_{_{R\rightarrow V}}\left(\frac{d_{r}}{N},\sigma_{n}^{2}\right)$.
Thus, we introduce a \textit{repetition to user load ratio }(RUR)\textit{
$\gamma_{\mathrm{RUR}}$} which is defined by 
\[
\gamma_{_{\mathrm{RUR}}}=\frac{d_{r}}{N}\,.
\]
According to the proof in Appendix B, we show that 
\begin{equation}
\underset{\underset{\frac{d_{r}}{N}\rightarrow\gamma_{_{\mathrm{RUR}}}}{N\rightarrow\infty}}{\mathrm{lim}}\:\mu_{_{R\rightarrow V}}\left(N,d_{r},\sigma_{n}^{2}\right)=\mu_{_{R\rightarrow V}}\left(\gamma_{_{\mathrm{RUR}}},\sigma_{n}^{2}\right).\label{eq:RUR}
\end{equation}
Empirically, the approximation of the ``channel input'' to the LDPC
decoder, expressed as
\begin{equation}
\mu_{_{R\rightarrow V}}\left(N,d_{r},\sigma_{n}^{2}\right)\approx\mu_{_{R\rightarrow V}}\left(\gamma_{_{\mathrm{RUR}}},\sigma_{n}^{2}\right),\label{eq:RepUserRatioAppro}
\end{equation}
is quite accurate also for small number of users through the EXIT
analysis using \eqref{eq:MUDRepConvMu} (EXIT curves not shown in
the paper, however, BER simulation results are shown in Fig.~\ref{fig:RepUserLoadSumrate}). 

This feature is particularly important in terms of adaptiveness to
the user load $N$. For instance, if the number of active users decreases
or increases, the LDPC decoder ``sees'' a better or worse channel,
respectively, which drifts the system operating point away from capacity\footnote{When the channel becomes better, the code rate of the LDPC code can
be higher.} or the system collapses\footnote{The optimized LDPC code at certain SNR can not operate when the real
SNR becomes smaller.}. If the approximation in \eqref{eq:RepUserRatioAppro} holds, we
may only need to vary the $d_{\mathrm{r}}$ accordingly to maintain
the constant RUR $\gamma_{\mathrm{RUR}}$ so that the ``channel input''
to the LDPC decoder remains unchanged and subsequently the coding
part is decoupled from the multi-user front-end via the repetition
code component. 

\begin{figure}[tbh]
\begin{centering}
\begin{tikzpicture} [spy using outlines=
	{magnification=2, connect spies}]
\begin{axis}[
width=\linewidth,
height=\linewidth,
xmajorgrids,
yminorticks=true,
ymajorgrids,
yminorgrids,       
legend style={at={(0,1)}, anchor=north west, legend cell align=left,align=left,draw=white!15!black, font=\footnotesize},
legend columns=2, 
legend style={
            /tikz/column 2/.style={
                column sep=5pt,
            },
},
legend entries={$d_{r}=1$,
                $d_{r}=2$,
                $d_{r}=3$,
                $d_{r}=4$,
                $d_{r}=6$,
                $d_{r}=8$,
                fix $\gamma_{\mathrm{RUR}}=\frac{1}{8}$,
                sim. BER<$10^{-4}$},
xlabel={No. of users $N$},
ylabel={Gap to GMAC capacity in dB},
xtick={8,16,24,32,40,48,56,64},
mark size=1.5pt,
xmin=6,
xmax=66,
ymin=1,
ymax=3.5
]	
\addlegendimage{mittelblau,mark=x}
\addlegendimage{rot,mark=o}
\addlegendimage{pink,mark=+}
\addlegendimage{apfelgruen,mark=square}
\addlegendimage{dunkelgrau,mark=triangle}
\addlegendimage{lila,mark=diamond}
\addlegendimage{orange, dashed, mark=none}
\addlegendimage{only marks, mark size=4pt, mark=pentagon*, color=myorgange}

\addplot[color=mittelblau, very thick,each nth point={1},mark=x] table [x index=0,y expr=\thisrowno{1}+0.365] {tikz/Data/GA_rep_user_rep1.dat};

\addplot[color= rot, very thick,each nth point={1},mark=o] table [x index=0,y expr=\thisrowno{1}+0.365] {tikz/Data/GA_rep_user_rep2.dat};

\addplot[color= pink, very thick,each nth point={1},mark=+] table [x index=0,y expr=\thisrowno{1}+0.365] {tikz/Data/GA_rep_user_rep3.dat};

\addplot[color= apfelgruen, very thick,each nth point={1},mark=square] table [x index=0,y expr=\thisrowno{1}+0.365] {tikz/Data/GA_rep_user_rep4.dat};

\addplot[color= dunkelgrau, very thick,each nth point={1},mark=triangle] table [x index=0,y expr=\thisrowno{1}+0.365] {tikz/Data/GA_rep_user_rep6.dat};

\addplot[color= lila, very thick,each nth point={1},mark=diamond] table [x index=0,y expr=\thisrowno{1}+0.365] {tikz/Data/GA_rep_user_rep8.dat};
\addplot[color= orange, very thick, dashed,each nth point={1}] table [x index=0, y expr=\thisrowno{1}+0.365] {tikz/Data/GA_rep_user_D.dat};

\node[color=mittelblau,regular polygon,regular polygon sides=5, inner sep=1pt,minimum size=10pt,draw,fill] at (axis cs:8,1.445) {};
\node[color= rot,regular polygon,regular polygon sides=5, inner sep=1pt, minimum size=10pt,draw,fill] at (axis cs:16,1.445) {};
\node[color= pink,regular polygon,regular polygon sides=5, inner sep=1pt, minimum size=10pt,draw,fill] at (axis cs:25,1.265) {};
\node[color= apfelgruen,regular polygon,regular polygon sides=5,  inner sep=1pt,minimum size=10pt,draw,fill] at (axis cs:33,1.315) {};
\node[color= dunkelgrau,regular polygon,regular polygon sides=5, inner sep=1pt, minimum size=10pt,draw,fill] at (axis cs:50,1.265) {};
\node[color= lila,regular polygon,regular polygon sides=5,  inner sep=1pt,minimum size=10pt, draw,fill] at (axis cs:64,1.445) {};


\end{axis}

\end{tikzpicture}
\par\end{centering}
\caption{GA-density evolution-based $\frac{E_{b}}{N_{0}}$-gap to GMAC capacity
for varying $\gamma_{_{\mathrm{RUR}}}$ with the \textbf{same} LDPC
code optimized for $d_{r}=4$, $N=32$ users and $\gamma_{s,\mathrm{mu}}=\unit[0]{dB}$;
the dashed line depicts the operation points with adaptive repetition
factor according to the user load by fixing the RUR to a constant
$\gamma_{_{\mathrm{RUR}}}=\frac{1}{8}$; pentagons denote actual simulation
results at the BER of $10^{-4}$; Note that the \textbf{same} $\mathbf{H}$-matrix
is used for all simulations.\label{fig:RepUserLoadThreshold}}
\end{figure}
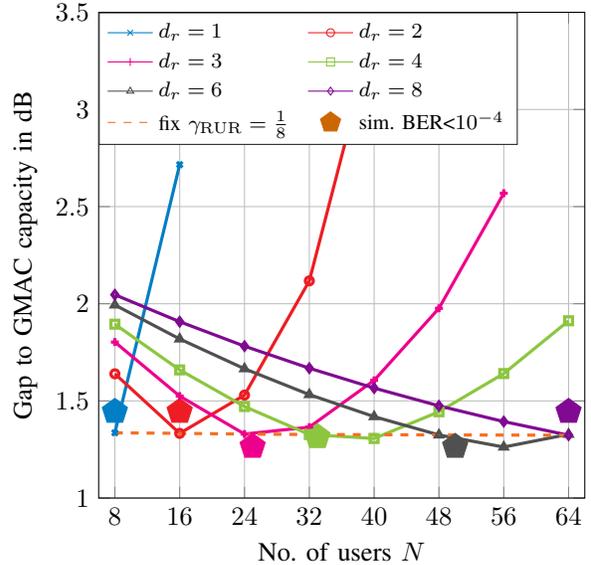

\begin{figure}[t]
\begin{centering}
\begin{tikzpicture} 
\begin{axis}[
width=\linewidth,
height=\linewidth,
legend pos=north west,        
xlabel={Repetition factor $d_{r}$},
ylabel={No. of users $N$},
mark size=10pt,
xmin=0.5,
xmax=8.5,
xtick={1,2,3,4,5,6,7,8},
ymin=1,
ymax=70,
view={0}{90},
colorbar horizontal, 
colorbar style={
	  at={(0.5,1.03)},anchor=south,
	  xticklabel pos=upper,
            xticklabel=$10^{\pgfmathprintnumber{\tick}}$
	},
colormap name=jetlight,
title style={yshift=1cm},
enlargelimits=false,
axis on top,
title=BER
]	

\addplot [matrix plot*, point meta = explicit] table [x index=0,y index=2, meta expr=log10(\thisrowno{4})] {tikz/Data/Rep_user_flex_tot.dat};

\addplot [color= black, very thick, dashed] coordinates {(1, 8) (8,64)};

\end{axis}

\end{tikzpicture}
\par\end{centering}
\caption{BER comparison of IDMA system with different repetition factor $d_{r}$
and number of users $N$; the LDPC code optimized for $d_{r}=4$,
$N=32$ and $\gamma_{s,\mathrm{mu}}=\unit[0]{dB}$ remains fixed;
the dashed black curve denotes $\gamma_{_{\mathrm{RUR}}}=\frac{1}{8}$.
\label{fig:RepUserLoadSumrate}}
\end{figure}
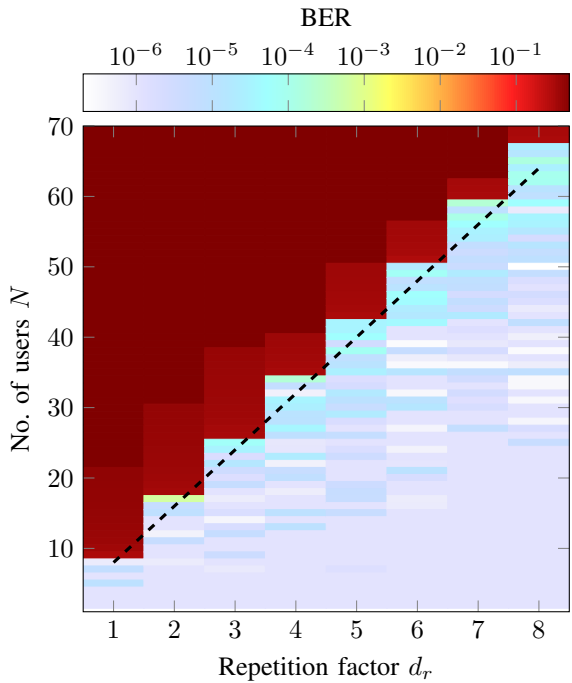

Fig. \ref{fig:RepUserLoadThreshold} shows the gap-to-GMAC capacity
where the LDPC code of rate $R_{c}=0.0975$ (optimized for $N=32$,
$d_{r}=4$ and $\gamma_{s,\mathrm{mu}}=\unit[0]{dB}$) is kept invariant.
The number of users $N$ and the repetition factor $d_{r}$ are varied.
Density evolution based on GA is used to estimate the decoding threshold
for each combination of $d_{r}$ and $N$. Obviously, the spectral
efficiency is quite invariant if the RUR $\gamma_{_{\mathrm{RUR}}}=\frac{1}{8}$
is kept constant. The BER simulation results are also included as
pentagons with corresponding colors in Fig. \ref{fig:RepUserLoadThreshold},
where the required SNR is obtained at the BER of $10^{-4}$. We note
that this adjustment of the repetition code does not incur any significant
loss of spectral efficiency and can be easily reconfigured at the
transmitter and receiver due to its simple structure and decoding
procedure. Furthermore, the simulated BERs at the multi user SNR $\gamma_{s,\mathrm{mu}}=\unit[0]{dB}$
are shown in Fig. \ref{fig:RepUserLoadSumrate} for $1\leq d_{r}\leq8$
and various $N$.
 The results consolidate our analytical and empirical expression
in \eqref{eq:RUR} and \eqref{eq:RepUserRatioAppro}. It also suggest
that the designed LDPC code for specific parameters can be easily
generalized and reused in systems with arbitrarily varying user load
by simply adjusting the repetition factor.

\section{\label{sec:Beyond-Equal-Power}Beyond Equal Power and Rate}

Different users' signals arrive at the base-station (BS) with different
power levels since the path loss between the BS and individual users
may be quite different. This fact indicates that the user rate shall
be adapted to its power level. We assume that each user transmits
at full power for higher data rates and propose two approaches in
the following to allow for multi-rate transmission, yet using the
same channel code that is optimized for the equal-power case.
\begin{enumerate}
\item \textit{Layer aggregation}. The higher power users may divide their
signals into several \textit{data layers or ``virtual users'', }each
with a smaller power and thus, the user with weakest power can only
transmit one data layer. This approach is straightforward, thus not
further discussed.
\item \textit{Repetition code adaptation (RCA). }The serially concatenated
repetition code can be further applied as a ``power equalizer''
in unequal receive (or transmit) power cases, or, more precisely,
as an ``energy equalizer''. Users with higher power can reduce the
repetition factor $d_{r}$ and users with lower power shall increase
the repetition factor. Let $P_{i}$ and $d_{r,i}$ denote the signal
power and repetition factor of the user $i$, we show that the ``channel
input'' to the $i$th LDPC decoder can be written as
\begin{equation}
\mu_{_{R\rightarrow V}}=\frac{4\gamma_{_{\mathrm{RCA}}}}{\sigma_{n}^{2}+\phi\left(\kappa\left(\gamma_{_{\mathrm{RCA}}}\right)\right)}\label{eq:RCApprox}
\end{equation}
in the asymptotic case of $P_{i}\rightarrow0$ while $P_{i}\cdot d_{r,i}\rightarrow\gamma_{_{\mathrm{RCA}}}=\mathrm{const.}$,
where $\kappa\left(\gamma_{_{\mathrm{RCA}}}\right)$ is a constant
depending on the feedback from the channel decoders and the value
of $\gamma_{_{\mathrm{RCA}}}$. A detailed derivation is given in
Appendix C.
\end{enumerate}
For simplicity of numerical simulation, we consider the following
unequal power IDMA system 
\[
y=\sqrt{P_{\mathrm{st}}}{\displaystyle \sum_{i=1}^{N/2}}\widetilde{x}_{i}+\sqrt{P_{\mathrm{wk}}}{\displaystyle \sum_{j=\frac{N}{2}+1}^{N}}\widetilde{x}_{j}+n
\]
where half of the users $\frac{N}{2}$ (without loss of generality,
$N$ is assumed to be even) exhibit a higher power than the other
half number of users, i.e., $P_{\mathrm{st}}>P_{\mathrm{wk}}$. The
total power is normalized to one, that is, 
\[
P_{\mathrm{st}}+P_{\mathrm{wk}}=\frac{2}{N}
\]
and we define a new quantity of ``power asymmetry'' as $P_{\mathrm{A}}=\frac{P_{\mathrm{st}}}{P_{\mathrm{wk}}}$.
The repetition code adaptation rule is given by 
\begin{align*}
\frac{d_{r,\mathrm{wk}}}{d_{r,\mathrm{st}}} & =P_{A}\\
\frac{1}{d_{r,\mathrm{wk}}}+\frac{1}{d_{r,\mathrm{st}}} & =\frac{2}{d_{r}}
\end{align*}
where $d_{r,\mathrm{wk}}$ and $d_{r,\mathrm{st}}$ denote the new
repetition factor for the strong and weak users, respectively. Notice
that if $d_{r,\mathrm{wk}}$ and $d_{r,\mathrm{st}}$ are not integer-valued,
irregular or quasi-regular repetition codes with the corresponding
average repetition factors shall be applied.

\begin{figure}[tbh]
\begin{centering}
\begin{tikzpicture} [spy using outlines=
	{magnification=2, connect spies}]
\begin{axis}[
width=\linewidth,
height=\linewidth,
xmajorgrids,
yminorticks=true,
ymajorgrids,
yminorgrids,
legend style={at={(0.5,1.25)},
      anchor=north,legend columns=2,legend cell align=left,align=left,draw=white!15!black, font=\footnotesize},
xlabel={$\frac{E_{b}}{N_{0}}$ in dB},
ylabel={BER},
ymode=log,
mark size=1.5pt,
xmin=-0.2,
xmax=1.3,
ymin=1e-4,
ymax=1
]	

\addplot[color= black, very thick,dashed ] coordinates {(-0.1, 1e-4) (-0.1,1e-2)};
\addlegendentry{GMAC capacity}



\addplot[color= rot,mark=o, very thick,each nth point={1}] table [x index=0, y index=1] {tikz/Data/BER_30user_UnequalP_Avg_Rep_4_snrs.dat};
\addlegendentry{Avg. all users, $P_{A}=3$dB}

\addplot[color= mittelblau, mark=x, very thick,each nth point={1}] table [x expr=\thisrowno{0}+0.2803, y index=1] {tikz/Data/Power_Gap_3dB_rep_4.dat};
\addlegendentry{REP Adap, $P_{A}=3$dB}

\addplot[color= dunkelgrau, mark=square, very thick,each nth point={1}] table [x index=0, y index=1] {tikz/Data/BER_30user_Rep_4_snrs.dat};
\addlegendentry{Equal Power, $P_{A}=0$dB}



\end{axis}

\end{tikzpicture}
\par\end{centering}
\caption{BER comparison of unequal-power IDMA system with the power asymmetry
of $P_{A}=\unit[3]{dB}$; the LDPC code is optimized for the equal-power
case with $d_{r}=4$, $N=32$.\label{fig:BERunequalPower}}
\end{figure}
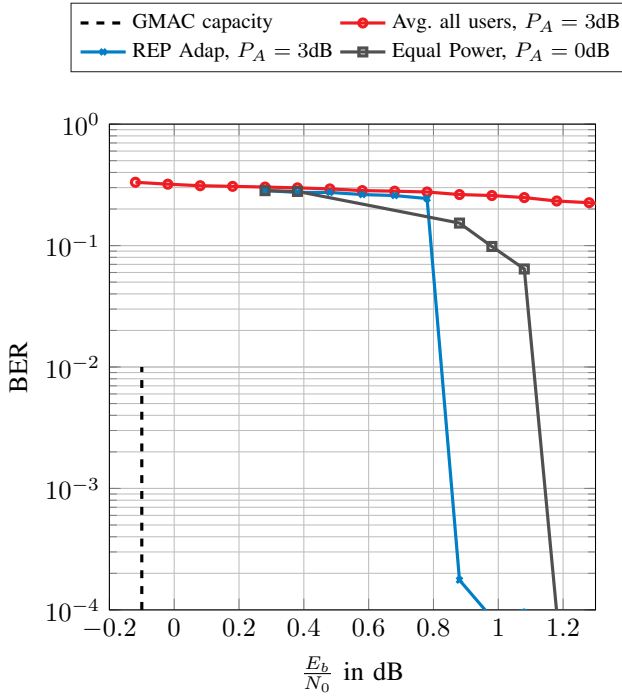

In Fig. \ref{fig:BERunequalPower}, 
we show the BER performance of an unequal-power IDMA system with the
power asymmetry between strong users and weak users of $P_{A}=\unit[3]{dB}$.
The LDPC code remains unchanged ($R_{c}=0.125$ and $d_{r}=4$). As
the LDPC code is optimized for the equal power case, the average performance
becomes worse since the weak users require higher SNRs. If the RCA\textit{
}is applied where the strong users employ a repetition factor of $d_{r,\mathrm{st}}=3$
and the weak users employ the new repetition factor of $d_{r,\mathrm{wk}}=6$
instead of $d_{r}=4$, then a similar performance can be achieved
compared to the equal power case, as the adapted repetition rate acts
as a ``power equalizer''. The extension to other cases of asymmetric
power distribution is straightforward.

\section{\label{sec:Conclusion}Conclusion}

We have proposed a flexible and simple coding scheme to achieve near-capacity
performance in a dynamic multiple access channel (MAC) system with
both a varying number of users as well as varying transmit (or receive)
power levels. In particular, a joint optimization of serially concatenated
LDPC and repetition code is considered in a low-complexity IDMA system.
 We show that near-capacity performance can be maintained with the
same LDPC code regardless of the number of users and signal power
levels. This universality of the proposed scheme is made possible
by the repetition code acting as both an interference equalizer as
well as power equalizer, which is of particularly practical interest.
The same LDPC parity check matrix allows to cover a remarkably wide
range of scenarios with respect to number of users and transmit powers,
making this highly flexible and simple NOMA scheme an attractive candidate
for future wireless communication systems.

\appendices{}

\section{\label{subsec:Proof-of-1}Proof of \eqref{eq:StabCond-1}}

The stability condition requires that the message $\mu$ approaches
infinity as the number of iterations goes to infinity. Assume that
a very large $\mu_{l}$ is reached after a finite number of iterations
and let $\tilde{\mu}_{l}$ denote 
\[
\tilde{\mu}_{l}=\mu_{l}+\frac{4}{N\sigma_{n}^{2}+\left(N-1\right)\cdot\phi\left(2\cdot\mu_{l}\right)}.
\]
The left term of \eqref{eq:StabCond-1} is lower-bounded with Taylor-series
by 
\begin{align*}
\mathrm{left\,term\,of\,\eqref{eq:StabCond-1}} & \geq1-\lambda_{2}\phi'\left(\tilde{\mu}_{l}\right)\cdot\Delta\mu\\
 & +\frac{8\lambda_{2}\phi'\left(2\mu_{l}\right)}{\left(N\sigma_{n}^{2}+\left(N-1\right)\cdot\phi\left(2\cdot\mu_{l}\right)\right)^{2}}\Delta\mu
\end{align*}
and when $\mu_{l}\rightarrow\infty$, we have $\tilde{\mu}_{l}\rightarrow\mu_{l}+\frac{4}{N\sigma_{n}^{2}}$
and the left term can be written as 
\[
1-\underset{\mu_{l}\rightarrow\infty}{\mathrm{lim}}\lambda_{2}\phi'\left(\mu_{l}+\underset{\mu_{\mathrm{ch}}}{\underbrace{\frac{4}{N\sigma_{n}^{2}}}}\right)\cdot\Delta\mu.
\]
Similarly, the right term of \eqref{eq:StabCond-1} is approximated
by 
\[
1-\frac{1}{d_{c}-1}\underset{\mu_{l}\rightarrow\infty}{\mathrm{lim}}\phi'\left(\mu_{l}\right)\cdot\Delta\mu.
\]
Therefore, we obtain 
\begin{align*}
\lambda_{2} & \leq\frac{1}{d_{c}-1}\cdot\underset{\mu_{l}\rightarrow\infty}{\mathrm{lim}}\frac{\phi'\left(\mu_{l}\right)}{\phi'\left(\mu_{l}+\mu_{\mathrm{ch}}\right)}\\
 & =\frac{e^{\frac{\mu_{\mathrm{ch}}}{4}}}{d_{c}-1}
\end{align*}
We note that $\phi\left(\mu\right)\approx\sqrt{\frac{\pi}{\mu}}e^{-\frac{\mu}{4}}$
is a tight approximation when $\mu$ is large.

\section{Proof of \eqref{eq:RUR}}

It is sufficient to show that the extrinsic output of the MUD+REP
remains the same for an arbitrary a priori information from the LDPC
decoder, provided that the RUR is constant, i.e., $\gamma_{_{\mathrm{RUR}}}=\mathrm{const.}$
Let $I_{R\leftarrow D}^{A}$ denote the a priori information from
the LDPC decoder to the repetition decoder. The extrinsic message
from repetition code to the MUD can be written as 
\[
I_{R\rightarrow M}^{E}=J\left(\left(d_{r}-1\right)\cdot J^{-1}\left(I_{MUD}^{E}\right)+J^{-1}\left(I_{R\leftarrow D}^{A}\right)\right)
\]
where $I_{MUD}^{E}$ denotes the extrinsic message from MUD to REP,
given in \eqref{eq:MUDextEXIT} with $I_{MUD}^{A}=I_{R\rightarrow M}^{E}$.
After sufficient iterations between MUD and REP, the intersection
of EXIT functions of MUD and REP can be written as 
\begin{align*}
J\left(\frac{J^{-1}\left(I_{R\rightarrow M}^{E}\right)-J^{-1}\left(I_{R\leftarrow D}^{A}\right)}{d_{r}-1}\right)=\\
J\left(\frac{4}{N\sigma_{n}^{2}+\left(N-1\right)\cdot\phi\left(J^{-1}\left(I_{R\rightarrow M}^{E}\right)\right)}\right).
\end{align*}
With some mathematical manipulation, we obtain 
\begin{equation}
\kappa\sigma_{n}^{2}+\left(1-\frac{1}{N}\right)\phi\left(\kappa\right)\cdot\left(\kappa-\nu_{D}\right)=\nu_{D}\sigma_{n}^{2}+4\frac{d_{r}-1}{N}\label{eq:MUDREPConvLDPC}
\end{equation}
where $\kappa=J^{-1}\left(I_{R\rightarrow M}^{E}\right)$ denotes
the converged message between MUD and REP for a given a priori information
from LDPC decoder $\nu_{D}=J^{-1}\left(I_{R\leftarrow D}^{A}\right)$. 

Let 
\begin{equation}
N\rightarrow\infty\:\mathrm{while}\,\frac{d_{r}}{N}\rightarrow\gamma_{_{\mathrm{RUR}}}\mathrm{=const.}\label{eq:AsymCond}
\end{equation}
Thereby, \eqref{eq:MUDREPConvLDPC} can be written as 
\[
\kappa\sigma_{n}^{2}+\phi\left(\kappa\right)\cdot\left(\kappa-\nu_{D}\right)=\nu_{D}\sigma_{n}^{2}+4\gamma_{_{\mathrm{RUR}}}.
\]
Clearly, the solution of $\kappa\left(\gamma_{_{\mathrm{RUR}}}\right)$
to the above equation merely depends on $\gamma_{_{\mathrm{RUR}}}$,
$\sigma_{n}^{2}$ and $\nu_{D}$. Therefore, the message from REP
to LDPC decoder for next iteration is given by 
\[
\mu_{_{R\rightarrow V}}=\frac{4d_{r}}{N\sigma_{n}^{2}+\left(N-1\right)\cdot\phi\left(\kappa\right)}\overset{\eqref{eq:AsymCond}}{=}\gamma_{_{\mathrm{RUR}}}\frac{4}{\sigma_{n}^{2}+\phi\left(\kappa\right)}.
\]

\section{Proof of \eqref{eq:RCApprox}}

We first assume that the messages passed from REP to MUD are of the
same mean value among all users, denoted by $\mu_{_{R\rightarrow M,i}}=J^{-1}\left(I_{R\rightarrow M}^{E}\right),\,\forall i$.
Following the same approach in Appendix B, the intersection of EXIT
functions of MUD and REP for the $i$th user can be written as 
\begin{align*}
J\left(\frac{J^{-1}\left(I_{R\rightarrow M}^{E}\right)-J^{-1}\left(I_{R\leftarrow D}^{A}\right)}{d_{r,i}-1}\right)=\\
J\left(\frac{4P_{i}}{\sigma_{n}^{2}+{\textstyle {\displaystyle \sum_{j=1,j\ne i}^{N}}}P_{j}\cdot\phi\left(J^{-1}\left(I_{R\rightarrow M}^{E}\right)\right)}\right).
\end{align*}
With some mathematical manipulation and let the total power be normalized
to 1, i.e., ${\displaystyle \sum_{i=1}^{N}}P_{i}=1$, we obtain 
\begin{equation}
\kappa\sigma_{n}^{2}+\left(1-P_{i}\right)\phi\left(\kappa\right)\cdot\left(\kappa-\nu_{D}\right)=\nu_{D}\sigma_{n}^{2}+4P_{i}\left(d_{r,i}-1\right)\label{eq:MUDREPConvLDPC-1}
\end{equation}
where $\kappa=J^{-1}\left(I_{R\rightarrow M}^{E}\right)$ denotes
the converged message between MUD and REP for a given a priori information
from LDPC decoder $\nu_{D}=J^{-1}\left(I_{R\leftarrow D}^{A}\right)$.
Considering the following asymptotic 
\begin{equation}
P_{i}\rightarrow0,\,P_{i}d_{r,i}\rightarrow\gamma_{_{\mathrm{RCA}}}\mathrm{=const.,\forall}i\label{eq:AsymCond-1}
\end{equation}
we can re-write \eqref{eq:MUDREPConvLDPC-1} as 
\begin{equation}
\kappa\sigma_{n}^{2}+\phi\left(\kappa\right)\cdot\left(\kappa-\nu_{D}\right)=\nu_{D}\sigma_{n}^{2}+4\gamma_{_{\mathrm{RCA}}}.\label{eq:RCAequ}
\end{equation}
Clearly, the solution of $\kappa\left(\gamma_{_{\mathrm{RCA}}}\right)$
to the above equation merely depends on $\gamma_{_{\mathrm{RCA}}}$.
Therefore, the message from REP to LDPC decoder for next iteration
is given by 
\begin{equation}
\mu_{_{R\rightarrow V}}=\frac{4P_{i}d_{r,i}}{\sigma_{n}^{2}+\left(1-P_{i}\right)\cdot\phi\left(\kappa\right)}\overset{\eqref{eq:AsymCond-1}}{=}\gamma_{_{\mathrm{RCA}}}\frac{4}{\sigma_{n}^{2}+\phi\left(\kappa\right)}.\label{eq:UnequalPowerRtoVSame}
\end{equation}
Next, we prove that the messages passed from REP to MUD are of the
same mean value among all users, e.g., $\mu_{_{R\rightarrow M,i}}=J^{-1}\left(I_{R\rightarrow M}^{E}\right),\,\forall i$.
According to \eqref{eq:RepToMUD}, the message mean from REP to MUD
of user $i$ can be written as 
\begin{align}
\mu{}_{_{R\rightarrow M,i}} & =\left(d_{r,i}-1\right)\frac{4P_{i}}{\sigma_{n}^{2}+\left(1-P_{i}\right)\cdot\phi\left(\kappa\right)}+\nu_{D}\label{eq:RtoMSameUnequal}\\
 & \overset{\eqref{eq:AsymCond-1}}{=}\mu_{_{R\rightarrow V}}+\nu_{D}.
\end{align}
Apparently, the message mean is the same for all users as long as
$\mu_{_{R\rightarrow V}}$ and $\nu_{D}$ remain unchanged. We have
proved that $\mu_{_{R\rightarrow V}}$ remains unchanged with the
proposed RCA and $\nu_{D}$ remains unchanged since we apply the same
LDPC code for all users.\balance

\bibliographystyle{IEEEtran}
\bibliography{bibliography}

\end{document}